\documentclass[prd,preprint,superscriptaddress,showpacs,byrevtex]{revtex4}
\usepackage{bm}
\def\fsl#1{\setbox0=\hbox{$#1$}           % set a box for #1
   \dimen0=\wd0                                 % and get its size
   \setbox1=\hbox{/} \dimen1=\wd1               % get size of /
   \ifdim\dimen0>\dimen1                        % #1 is bigger
      \rlap{\hbox to \dimen0{\hfil/\hfil}}      % so center / in box
      #1                                        % and print #1
   \else                                        % / is bigger
      \rlap{\hbox to \dimen1{\hfil$#1$\hfil}}   % so center #1
      /                                         % and print /
   \fi}                                         %
\newcommand{\be}{\begin{equation}}
\newcommand{\ee}{\end{equation}}
\newcommand{\bea}{\begin{eqnarray}}
\newcommand{\eea}{\end{eqnarray}}
\newcommand{\beq}{\begin{equation}}
\newcommand{\eeq}{\end{equation}}
\newcommand{\beqs}{\begin{eqnarray}}
\newcommand{\eeqs}{\end{eqnarray}}

\begin{document}
\title{ General Form of the Color Potential Produced by Color Charges of the Quark }
\author{Gouranga C Nayak } \email{nayak@max2.physics.sunysb.edu}
\affiliation{1001 East 9th Street, \#A, Tucson, AZ 85719, USA}
%
%\date{\today}
%
\begin{abstract}
Constant electric charge $e$ satisfies the continuity equation $\partial_\mu j^{\mu }(x)= 0$ where
$j^\mu(x)$ is the current density of the electron. However, the Yang-Mills color current density
$j^{\mu a}(x)$ of the quark satisfies the equation $D_\mu[A] j^{\mu a}(x)= 0$ which is not a continuity equation
($\partial_\mu j^{\mu a}(x)\neq 0$) which implies that a color charge $q^a(t)$ of the quark is not constant but it is
time dependent where $a=1,2,...8$ are color indices. In this paper
we derive general form of color potential produced by color charges of the quark.
We find that the general form of the color potential produced by the color charges
of the quark at rest is given by
$\Phi^a(x) =A_0^a(t,{\bf x}) =\frac{q^b(t-\frac{r}{c})}{r}\left[\frac{{\rm exp}[g\int dr \frac{Q(t-\frac{r}{c})}{r}]
-1}{g \int dr \frac{Q(t-\frac{r}{c})}{r}}\right]_{ab}$ where $dr$ integration is an indefinite integration, ~~
$Q_{ab}(\tau_0)=f^{abd}q^d(\tau_0)$, ~~$r=|{\vec x}-{\vec X}(\tau_0)|$, ~~$\tau_0=t-\frac{r}{c}$
is the retarded time, ~~$c$ is the speed of light,
~~${\vec X}(\tau_0)$ is the position of the quark at the retarded time and the repeated color indices $b,d$(=1,2,...8) are summed.
For constant color charge $q^a$ we reproduce the Coulomb-like potential $\Phi^a(x)=\frac{q^a}{r}$
which is consistent with the Maxwell theory where constant electric charge $e$ produces the Coulomb potential
$\Phi(x)=\frac{e}{r}$.
\end{abstract}
\maketitle
\pagestyle{plain}
\pagenumbering{arabic}

\section{Introduction}

In the eighteenth century, Coulomb, in an impressive series of experiments
found that the static potential produced by an electric charge is inversely
proportional to distance. This is known as Coulomb's law. The atomic bound
state, such as hydrogen atom, is described by using Coulomb potential. However,
an exact form of the color potential produced by color charge is not known till now.
The exact form of color potential produced by color charges of the quark may
provide an insight to the question "why quarks are confined inside a (stable) proton".

Note that, in Maxwell theory fundamental electric charge $e$ of the electron is
constant. This constant electric charge $e$ satisfies the continuity equation $\partial_\mu j^\mu(x)=0$
where $j^\mu(x)$ is the current density of the electron. However, in Yang-Mills theory the Yang-Mills
color current density $j^{\mu a}(x)$ of the quark satisfies the equation $D_\mu[A]j^{\mu a}(x)=0$ which
is not the continuity equation ($\partial_\mu j^{\mu a}(x)\neq 0$) where
$D_\mu^{ab}[A]=\delta^{ab}\partial_\mu +gf^{acb}A_\mu^c(x)$ and $A^{\mu a}(x)$ is the Yang-Mills
potential (color potential). This implies that, unlike electric
charge $e$ of the electron which is constant, a color charge of the quark is not constant.

Also, it is important that the conserved color charges are not directly observable
-- only color representations -- because of the unbroken gauge invariance of QCD. Thus,
the concept of constant color charge seems unphysical. The asymptotic freedom \cite{asym}
which includes quantum effects (loop diagrams) predicts that the strong coupling $\alpha_s$
decreases at short distances and increases at long distances.

In quantum mechanics the charge density and probability density of the point particle can be obtained from
the quantum wave functions. Hence one finds that the charge density is more microscopic and may
depend on the space coordinate ${\vec x}$.
For example, consider charge and charge density in a volume $V=\int d^3{\vec x}$. The total charge in the volume
$V=\int d^3{\vec x}$ can be obtained from the corresponding charge density
after integrating over the entire (physically) allowed volume $V=\int d^3{\vec x}$. Hence one finds that
the total charge in the volume $V=\int d^3{\vec x}$ is independent of space coordinate ${\vec x}$.
If there is only one point charge in the entire (physically) allowed volume $V=\int d^3{\vec x}$ then one
finds that the charge of the point particle is independent of the space coordinate ${\vec x}$.

This implies that if the color charge of the quark is not constant then it has to be time dependent.

Since the color current density $j^{\mu a}(x)$ of a quark has eight color indices $a=1,2,...8$ one
finds that a quark has eight time dependent color charges $q^a(t)$ where $a=1,2,...8$. It is useful
to remember that the indices $i=1,2,3=$RED, BLUE, GREEN are not color charges of a quark but they
are color indices of the quark field $\psi_i(x)$.

We denote eight time dependent fundamental color charges of a quark by $q^a(t)$ where $a=1,2,...8$
are color indices. A color charge of a quark is flavor independent, {\it i.e.}, a color charge
$q^a(t)$ of the $u$ (up) quark is same as that of the $d$ (down), $S$ (strange), $C$ (charm), $B$
(bottom) or $t$ (top) quark.

It is interesting to note that lattice QCD \cite{wilson} can not predict the
form of the Yang-Mills potential (color potential) $A^{\mu a}(x)$ having color indices
$a=1,2,..8$ produced from color charge $q^a(t)$. The Wilson loop yields a gauge-invariant measure of the static
color singlet quark (effective) potential (energy). In Bohr's atomic model the Coulomb potential
provides the force of attraction between proton and electron to form (stable) hydrogen atom.
Similarly the color potential (Yang-Mills potential) produced by the color charges may provide the
force of attraction between quarks to form (stable) proton.

In the abelian gauge theory the Maxwell equation for electromagnetic potential $A^\mu(x)$ is given by
\bea
\partial_\mu F^{\mu \nu }(x)=j^{\nu }(x)
\label{1aa}
\eea
where
\bea
F^{\mu \nu}(x)=\partial^\mu A^\nu(x) - \partial^\nu A^\mu(x).
\label{fmn}
\eea
The Dirac equation of the electron is given by
\bea
[i\gamma^\mu \partial_\mu +e\gamma^\mu A_\mu(x)-m]\psi(x)=0.
\label{dea}
\eea
The current density of the electron
\bea
j^\mu(x)= e{\bar \psi}(x)\gamma^\mu \psi(x)
\label{aaa}
\eea
obeys the continuity equation
\bea
\partial_\mu j^{\mu}(x)=0.
\label{abaa}
\eea
Experimentally, it was discovered by Millikan that the fundamental
electric charge $e$ is a constant. From the continuity
equation (\ref{abaa}) in Maxwell theory we find that the
electric charge $e$ is constant which is consistent with the experimental
finding of Millikan. Note that we have neglected the vacuum polarization
effects. When the vacuum polarization effect is included the effective electric
charge $e_{eff}$ increases as $r$ decreases.

The situation in Yang-Mills theory is different, as we will see below.

The Yang-Mills equation for the non-abelian potential (Yang-Mills potential)
$A^{\mu a}(x)$ is given by \cite{yang,mutta}
\bea
D_\mu[A]F^{\mu \nu a}(x)=j^{\nu a}(x)
\label{1a}
\eea
where
\bea
D_\mu^{ab}[A]=\delta^{ab}\partial_\mu +gf^{acb}A_\mu^c(x)
\label{dm}
\eea
and
\bea
F^{\mu \nu a}(x)=\partial^\mu A^{\nu a}(x) - \partial^\nu A^{\mu a}(x)+gf^{abc}A^{\mu b}(x)A^{\nu c}(x).
\label{fmna}
\eea
The Dirac equation of the quark is given by
\bea
[\delta_{ij}(i\gamma^\mu \partial_\mu -m)+gT^a_{ij}\gamma^\mu A_\mu^a(x)]\psi_j(x)=0.
\label{de}
\eea
The Yang-Mills color current density of the quark
\bea
j^{\mu a}(x)= g{\bar \psi}_i(x)T^a_{ij}\gamma^\mu \psi_j(x)
\label{aa}
\eea
obeys the equation
\bea
D_\mu[A]j^{\mu a}(x)=0.
\label{ab}
\eea

As discussed above since eq. (\ref{ab}) is not a continuity equation ($\partial_\mu j^{\mu a}(x)\neq 0$),
a color charge $q^a(t)$ of the quark is time dependent. For earlier works on classical Yang-Mills theory,
see \cite{wuyang,wuyang1,mandula,all}.

In this paper we derive general form of color potential produced by the color
charges of the quark.

Note that the Yang-Mills theory was developed in analogy to the procedure in electromagnetic
theory (Maxwell theory) by extending the U(1) gauge group to the SU(3) gauge group appropriately
\cite{yang}. For example in analogy to $D^\mu \psi = (\partial^\mu -ie A^\mu)\psi$ in
electromagnetic theory (Maxwell theory) all derivatives of $\psi$ in Yang-Mills theory appear
in the combination $D^\mu \psi = (\partial^\mu -i\epsilon B^\mu)\psi$ where $B^\mu =T^a A^{\mu a}$
\cite{yang}. Similarly, in analogy to commutator equation $[D^\mu,~D^\nu]=ieF^{\mu \nu}$ in
electromagnetic theory (Maxwell theory) which predicts the relation between $F^{\mu \nu}(x)$
and $A^\mu(x)$ as given by eq. (\ref{fmn}) one finds that in Yang-Mills theory the commutator equation
$[D^\mu,~D^\nu]=-igT^aF^{\mu \nu a}$ predicts the relation between $F^{\mu \nu a}(x)$
and $A^{\mu a}(x)$ as given by eq. (\ref{fmna}) \cite{mutta,yang}. Similarly, in analogy to the
electromagnetic field Lagrangian density ${\cal L}_{\rm EM}=-\frac{1}{4} F_{\mu \nu}F^{\mu \nu}$ in
electromagnetic theory (Maxwell theory) where $F^{\mu \nu}(x)$ is given by eq. (\ref{fmn}) one writes down
the Yang-Mills field Lagrangian density ${\cal L}_{\rm YM}=-\frac{1}{4} F_{\mu \nu}^aF^{\mu \nu a}$ in
Yang-Mills theory where $F^{\mu \nu a}(x)$ is given by eq. (\ref{fmna}) \cite{yang}.
Hence in order to get an idea of the general form of the color potential (Yang-Mills potential)
produced by the color charges of the quark we make analogy between electromagnetic theory
(Maxwell theory) and Yang-Mills theory by transforming pure gauge classical potential.
Note that in Maxwell theory an electric charge moving at speed of light produces U(1) pure gauge potential,
see \cite{sterman}. In analogy to Maxwell theory one finds in Yang-Mills theory
that the color charges moving at speed of light produce SU(3) pure gauge potential
(see section IVC). The form of the
physical electromagnetic potential (Lienard-Wiechert potential or Coulomb potential) produced by the
electric charge $e$ of the electron can be obtained from the form of U(1) pure gauge potential produced
by the electron. In analogy to the electromagnetic case we find that in Yang-Mills theory
the form of physical color potential (Yang-Mills potential) produced by the color charges $q^a(t)$ of the
quark can be obtained from the form of SU(3) pure gauge potential produced by the quark.

The above analogy between electromagnetic theory (Maxwell theory) and Yang-Mills theory to
obtain the form of the physical potential from the form of the pure gauge potential can
be proved by using Lorentz transformation.
First of all we note that an electron has non-zero
mass (even if very small) and hence it can not
travel at speed of light $v=c$. The highest speed of the electron can be $v\sim c$ (which is
arbitrarily close to the speed of light). Hence the electron in uniform motion at this highest speed
$v \sim c$ produces U(1) (approximate) pure gauge potential \cite{sterman}.
We call this as (approximate) pure gauge potential because the highest speed of the electron can be
$v\sim c$ (if $v=c$ then a charge produces exact pure gauge potential \cite{sterman}), see section
IIIA for details. The exact form
of the Coulomb potential produced by the electron at rest is obtained from the form of U(1) (approximate)
pure gauge potential by putting $v=0$ in the U(1) (approximate) pure gauge potential.
Since the highest speed of the electron can be $v\sim c$ (but
not $v = c$), this can also be shown by using Lorentz transformations
(see section V and appendix B). Similarly in Yang-Mills theory a quark has non-zero mass
(even if the mass of the light quark is very small) and hence a quark can not
travel at speed of light $v=c$. The highest speed of a quark can be $v\sim c$ (which is
arbitrarily close to the speed of light). Hence the quark in uniform motion at this highest speed
$v \sim c$ produces SU(3) (approximate) pure gauge potential (see section IVC). The exact form of the color
potential produced by the quark at rest is obtained from the form of SU(3) (approximate) pure
gauge potential by putting $v=0$ in the
SU(3) (approximate) pure gauge potential. Since the highest speed of a quark can be $v\sim c$ (but not $v=c$),
this can also be shown by using Lorentz transformations (see section V).

We find that the general form of the color potential $\Phi^a(x) =A_0^a(t,{\bf x})$ produced by
the color charges of the quark at rest is given by
\bea
\Phi^a(x)=A_0^a(t,{\bf x}) =\frac{q^b(t-\frac{r}{c})}{r}\left[\frac{{\rm exp}[g\int dr \frac{Q(t-\frac{r}{c})}{r}]
-1}{g \int dr \frac{Q(t-\frac{r}{c})}{r}}\right]_{ab}
\label{upb}
\eea
where $dr$ integration is an indefinite integration, $a$=1,2,..8 are color indices,
$c$ is the speed of light, $\tau_0=t-\frac{r}{c}$ is the retarded time,
\bea
r=|{\vec x}-{\vec X}(\tau_0)|,~~~~~~~~~~~~~~~~Q_{ab}(\tau_0)=f^{abd}q^d(\tau_0),
\label{ivq3}
\eea
${\vec X}(\tau_0)$ is the position of the quark at the retarded time, and the repeated color indices $b,d$(=1,2,...8)
are summed. The non-zero values of $f^{abd}$ are given by
\bea
\begin{array}{cccccccccc}
{\underline{abd}}~&~123~&~147~&~156~&~246~&~257~&~345~&~367~&~458~&~678\\
{\underline {f^{abd}}}~&~ 1~&~\frac{1}{2}~&~-\frac{1}{2}~&~\frac{1}{2}~&~\frac{1}{2}~&~\frac{1}{2}~&~-\frac{1}{2}~&~\frac{\sqrt{3}}{2}~&~\frac{\sqrt{3}}{2}
\end{array}
\label{fabclist}
\eea
where $f^{abd}=-f^{bad}=-f^{adb}$. All other values of $f^{abd}$ are zero.

For constant color charge $q^a$, eq. (\ref{upb}) reproduces the Coulomb-like
potential $\Phi^a(x)=\frac{q^a}{r}$, which is consistent with the Maxwell
theory where constant electric charge $e$ produces Coulomb potential $\Phi(x)=\frac{e}{r}$.

Note that the exact form of color potential produced by color charges of the quark may
provide an insight to the question "why quarks are confined inside a (stable) proton".
Hence in order to find the exact form of the color potential produced by the color charges of the quark we
need to find the exact form of eight time dependent fundamental color charges $q^a(t)$ of the quark where
$a=1,2,...8$ are color indices. The general form of the time dependent fundamental color charge $q^a(t)$
of the quark in Yang-Mills theory in SU(3) is defined in \cite{nayak2} which we have briefly described
in section VI.

Note that at short distances, the quantum effects via
vacuum polarization can be incorporated in to the color charge $q^a(\tau_0)$.
The coupling $g$ becomes very small at short distances due to
quantum effects/loop diagrams. Hence at short distances where quantum effects are
important we find from eq. (\ref{upb})
\bea
\Phi^a(x) \rightarrow \frac{q^a(\tau_0)}{r},~~~~~~~~~~{\rm as }~~~~~~~~~g\rightarrow 0
\label{short}
\eea
which implies that general form of color potential produced by the color charges of the
quark reduces to Coulomb-like potential
at short distances. This seems to be in agreement with lattice QCD predictions where the effective
(color singlet) static confinement (effective) potential energy reduces to Coulomb form at short distances.

Note that the color potential $\Phi^a(x)$ in eq. (\ref{upb}) is not gauge invariant. The gauge
invariant quantity is chromo-electric field square $[\sum_{a=1}^8 {\vec E}^a(x) \cdot {\vec E}^a(x)]$.
The gauge invariant $[\sum_{a=1}^8 {\vec E}^a(x) \cdot {\vec E}^a(x)]$ can be obtained from eq. (\ref{upb}).

We will present a derivation of eq. (\ref{upb}) in this paper.

The paper is organized as follows. In section II we simplify all the infinite number
of non-commuting terms in the SU(3) pure gauge. In section III we derive the
expression of the U(1) pure gauge potential produced by the electron in Maxwell theory.
The general expression of the SU(3) pure gauge potential produced by the quark is
derived in section IV. In section V we derive general form of the color potential
$\Phi^a(x) =A_0^a(t,{\bf x})$ produced by the color charges of the quark at rest as given
by eq. (\ref{upb}). In section
VI we briefly describe the general form of the fundamental time dependent color charge
$q^a(t)$ of the quark in Yang-Mills theory in SU(3). Section VII contains conclusions.

\section{ Simplification of Infinite Number of Non-Commuting Terms in SU(3) Pure Gauge }

The SU(3) pure gauge potential $A^{\mu a}(x)$ is given by \cite{yang,mutta}
\bea
T^aA^{\mu a}(x) = \frac{1}{ig}[\partial^\mu U(x)]U^{-1}(x)
\label{pure}
\eea
where $T^a$ is the generator in the fundamental representation of SU(3) group (Gell-Mann matrices)
and
\bea
U(x)=e^{igT^a\omega^a(x)}.
\label{u}
\eea
The repeated indices $a$=1,2,...8 are summed. From eq. (\ref{u}) we find
\bea
&& [\partial^\mu U(x)]=[\partial^\mu \omega^d(x)] \frac{d}{d\omega^d(x)} [1+ igT^a \omega^a(x) + \frac{(ig)^2}{2!} T^aT^b \omega^a(x) \omega^b(x)+ \frac{(ig)^3}{3!} T^aT^bT^c \omega^a(x) \omega^b(x) \omega^c(x) \nonumber \\
&& +\frac{(ig)^4}{4!} T^aT^bT^cT^e \omega^a(x) \omega^b(x) \omega^c(x) \omega^e(x)+...].
\label{gb}
\eea
Since $T^a$ and $T^b$ do not commute with each other we find from eq. (\ref{gb})
\bea
&&[\partial^\mu U(x)] =[\partial^\mu \omega^d(x)] [igT^d + (ig)^2 T^dT^a \omega^a(x) + \frac{(ig)^3}{2!} T^dT^aT^b \omega^a(x) \omega^b(x)+\frac{(ig)^4}{3!} T^dT^aT^bT^c\omega^a(x) \nonumber \\
&&~ \omega^b(x) \omega^c(x)+.... \nonumber \\
&&+ \frac{(ig)^2}{2!} (T^aT^d-T^dT^a) \omega^a(x) + \frac{(ig)^3}{3!} (T^bT^dT^a+T^aT^bT^d-2T^dT^aT^b) \omega^a(x) \omega^b(x) \nonumber \\
&& +\frac{(ig)^4}{4!} (T^aT^bT^cT^d+T^bT^cT^dT^a+T^cT^dT^aT^b-3T^dT^aT^bT^c) \omega^a(x) \omega^b(x) \omega^c(x)+...].
\label{gc1s}
\eea
Note that the first infinite series in eq. (\ref{gc1s}) is $igT^dU$. Hence we find from eq. (\ref{gc1s})
\bea
&&[\partial^\mu U(x)] =[\partial^\mu \omega^d(x)] [igT^dU+ \frac{(ig)^2}{2!} (T^aT^d-T^dT^a) \omega^a(x) + \frac{(ig)^3}{3!} (T^bT^dT^a+T^aT^bT^d-2T^dT^aT^b) \nonumber \\
&&~ \omega^a(x) \omega^b(x)+\frac{(ig)^4}{4!} (T^aT^bT^cT^d+T^bT^cT^dT^a+T^cT^dT^aT^b-3T^dT^aT^bT^c) \omega^a(x) \omega^b(x) \omega^c(x)+...]. \nonumber \\
\label{gc}
\eea

By using the commutation relation \cite{mutta}
\bea
[T^a,~T^b]=if^{abc}T^c
\label{tab}
\eea
we find from eq. (\ref{gc})
\bea
&&[\partial^\mu U(x)] =[\partial^\mu \omega^d(x)] [igT^d U(x)+ \frac{(ig)^2}{2!} if^{ade} \omega^a T^e  [1+igT^b \omega^b(x)]+\frac{(ig)^3}{3!}if^{ade} \omega^a(x) if^{bej}\omega^b(x)T^j \nonumber \\
&& +\frac{(ig)^4}{4!} (if^{ade}T^eT^bT^c+if^{ade} T^bT^eT^c+if^{ade}T^cT^bT^e+if^{ade}T^eT^cT^b+if^{ade}T^bT^eT^c \nonumber \\
&& +if^{ade}T^eT^cT^b) \omega^a(x) \omega^b(x) \omega^c(x)+...].
\label{tc}
\eea
By repeated use of eq. (\ref{tab}) in (\ref{tc}) we finally obtain
\bea
&& [\partial^\mu U(x)]=[\partial^\mu \omega^d(x)] [igT^d U(x)+ \frac{(ig)^2}{2!} if^{ade} \omega^a(x) T^e  [1+igT^b \omega^b(x)]+\frac{(ig)^3}{3!}if^{ade} \omega^a(x) if^{bej}\omega^b(x)T^j \nonumber \\
&& +\frac{(ig)^4}{4!} (6if^{ade}T^eT^bT^c+4if^{ade}if^{bej}T^jT^c+if^{ade}if^{bej}if^{cjh}T^h) \omega^a(x) \omega^b(x) \omega^c(x)+...].
\label{td}
\eea
Simplifying eq. (\ref{td}) we find
\bea
&& [\partial^\mu U(x)] =[\partial^\mu \omega^d(x)] [igT^d U(x)+ \frac{(ig)^2}{2!} if^{ade} \omega^a(x) T^e  [1+igT^b \omega^b(x)+\frac{(ig)^2}{2!}T^b \omega^b(x)T^c\omega^c(x)]\nonumber \\
&& +\frac{(ig)^3}{3!}if^{ade} \omega^a(x) if^{bej}\omega^b(x)T^j  [1+igT^c\omega^c(x)]+\frac{(ig)^4}{4!} if^{ade}\omega^a(x)if^{bej} \omega^b(x)if^{cjh}\omega^c(x)T^h +...]. \nonumber \\
\eea
Generalizing this to infinity order we obtain
\bea
&& [\partial^\mu U(x)] =[\partial^\mu \omega^d(x)] [igT^d U(x)+ \frac{(ig)^2}{2!} if^{ade} \omega^a(x) T^e  [1+igT^b \omega^b(x)+\frac{(ig)^2}{2!}T^b \omega^b(x)T^c\omega^c(x)+....] \nonumber \\
&& +\frac{(ig)^3}{3!}if^{ade} \omega^a(x) if^{bej}\omega^b(x)T^j [1+igT^c\omega^c(x)+....]\nonumber \\
&&+\frac{(ig)^4}{4!} if^{ade}\omega^a(x)if^{bej} \omega^b(x)if^{cjh}\omega^c(x)T^h [1+....] +...].
\label{hhg}
\eea
Note that the expansion in each square bracket is $U(x)$, where $U(x)$ is given by eq. (\ref{u}). Hence we find from eq.
(\ref{hhg})
\bea
&&~\partial^\mu U(x)=[\partial^\mu \omega^d(x)] [igT^d U(x)+ \frac{(ig)^2}{2!} if^{ade} \omega^a(x) T^e U(x)
+\frac{(ig)^3}{3!}if^{ade} \omega^a(x) if^{bej}\omega^b(x)T^j U(x) \nonumber \\
&& +\frac{(ig)^4}{4!} if^{ade}\omega^a(x)if^{bej} \omega^b(x)if^{cjh}\omega^c(x)T^h U(x) +...].
\label{pl1}
\eea
Multiplying $U^{-1}(x)$ from right in eq. (\ref{pl1}) we find
\bea
&&~[\partial^\mu U(x)]U^{-1}(x)=[\partial^\mu \omega^d(x)] [igT^d + \frac{(ig)^2}{2!} if^{ade} \omega^a(x) T^e
+\frac{(ig)^3}{3!}if^{ade} \omega^a(x) if^{bej}\omega^b(x)T^j  \nonumber \\
&& +\frac{(ig)^4}{4!} if^{ade}\omega^a(x)if^{bej} \omega^b(x)if^{cjh}\omega^c(x)T^h +...].
\label{pla}
\eea
Simplifying eq. (\ref{pla}) we find
\bea
&&~\frac{1}{ig}[\partial^\mu U(x)]U^{-1}(x)=[\partial^\mu \omega^b(x)] [\delta^{ab} + \frac{ig}{2!} if^{bae} \omega^e(x)
+\frac{(ig)^2}{3!}if^{bej} \omega^j(x) if^{ead}\omega^d(x)  \nonumber \\
&& +\frac{(ig)^3}{4!} if^{beh}\omega^h(x)if^{ejd} \omega^d(x)if^{jac}\omega^c(x) +...]T^a.
\label{pla4}
\eea

Hence from eqs. (\ref{pla4}) and (\ref{pure}) we find
\bea
&&T^aA^{\mu a}(x)= [\partial^\mu \omega^b(x)] [\delta^{ab} + \frac{ig}{2!} if^{bae} \omega^e(x)
+\frac{(ig)^2}{3!}if^{bej} \omega^j(x) if^{ead}\omega^d(x)  \nonumber \\
&& +\frac{(ig)^3}{4!} if^{beh}\omega^h(x)if^{ejd} \omega^d(x)if^{jac}\omega^c(x) +...]T^a.
\label{pl}
\eea

From eq. (\ref{pl}) we find that the SU(3) pure gauge potential $A^{\mu a}(x)$ is given by
\bea
A^{\mu a} (x)=[\partial^\mu \omega^b(x)]~ [1 + \frac{g}{2!}M(x)
+\frac{g^2}{3!}M^2(x)+\frac{g^3}{4!} M^3(x)  +...]_{ab}
\label{54j}
\eea
where $M_{ab}(x)$ is given by
\bea
M_{ab}(x)=f^{abc}\omega^c(x).
\label{mab}
\eea

Hence we find from eq. (\ref{54j}) that the SU(3) pure gauge potential is given by
\bea
A^{\mu a} (x)=[\partial^\mu \omega^b(x)]~ [\frac{e^{gM(x)}-1}{gM(x)}]_{ab}
\label{su3pure}
\eea
where $M_{ab}(x)$ is given by eq. (\ref{mab}).

The abelian-like non-abelian pure gauge is given by
\bea
{\cal A}^{\mu a}(x) = \partial^\mu \omega^a(x).
\label{purea}
\eea
It can be seen that, unlike QED, the SU(3) pure gauge in QCD contains infinite number of higher order
terms in $\omega^a(x)$. Only the first term in eq. (\ref{su3pure}) or in eq. (\ref{54j}) corresponds
to the U(1) pure gauge in QED, see eq. (\ref{purea}).

\section{ Expression of the U(1) Pure Gauge Potential Produced by the Electron}

Consider the motion of an electron of electric charge $e$. Let ${X}^\mu(\tau)$
be the time-space position of the electron moving with the four-velocity
\bea
u^\mu(\tau) = \frac{d{X}^\mu(\tau)}{d\tau}.
\label{4va}
\eea
In classical electrodynamics the exact expression of the electromagnetic potential
(Lienard-Weichert potential)
\bea
A^\mu(x) = e \frac{u^\mu(\tau_0)}{u(\tau_0) \cdot (x-X(\tau_0))}
\label{1yj1}
\eea
produced by the constant electric charge $e$ of the electron in motion can
be derived by using the abelian electric current density
\bea
j^\mu(x)=\int d\tau ~e~u^\mu(\tau)~\delta^{(4)}(x-X(\tau))
\label{mxcz}
\eea
of the electron in the inhomogeneous wave equation \cite{jackson}
\bea
\partial^\nu \partial_\nu A^\mu(x)=j^\mu(x)
\label{mxpz}
\eea
by using $j^\mu(x)$ from eq. (\ref{mxcz}) in
\bea
{A}^{\mu }(x)=\int d^4x' D_r(x-x'){ j}^{\mu }(x')
\eea
where $D_r(x-x')$ is the retarded greens function.
In eq. (\ref{1yj1}) the $x^\mu=(ct,x,y,z)$ is the time-space position at which the electromagnetic field is
observed, $c$ is the speed of light and $\tau_0$ is defined by the light-cone condition
\bea
(x-X(\tau_0))^2=0
\label{lc}
\eea
and the retardation requirement
\bea
x_0 >X_0(\tau_0).
\label{rett}
\eea
From eqs. (\ref{lc}) and (\ref{rett}) one finds that the $\tau_0$ is
determined from the solution of the retarded condition
\bea
x_0-X_0(\tau_0) = |{\vec x} -{\vec X}(\tau_0)|.
\label{mx7}
\eea
Eq. (\ref{mx7}) [or eq. (\ref{lc})] implies that
\bea
\partial^\mu [(x-X(\tau_0))^\nu (x-X(\tau_0))_\nu] =0
\label{mununu}
\eea
which gives [by using eq. (\ref{4va})]
\bea
\partial^\mu \tau_0=\frac{x^\mu-X^\mu(\tau_0) }{u(\tau_0) \cdot (x-X(\tau_0))}.
\label{dtau0}
\eea

From eqs. (\ref{4va}), (\ref{mx7}) and  (\ref{dtau0}) one finds
\bea
\partial^\mu [u(\tau_0) \cdot (x-X(\tau_0))] =[\frac{[\dot{u}(\tau_0) \cdot (x-X(\tau_0))-c^2](x-X(\tau_0))^\mu }{u(\tau_0) \cdot (x-X(\tau_0))}+u^\mu(\tau_0)]
\label{dvu}
\eea
where
\bea
{\dot u}^\mu(\tau_0) = \frac{du^\mu(\tau)}{d\tau}|_{\tau=\tau_0}.
\label{ttp}
\eea
The electromagnetic potential $A^\mu(x)$ in eq. (\ref{1yj1}) is also
known as abelian potential or U(1) potential.
From eqs. (\ref{4va}), (\ref{mx7}), (\ref{dtau0}) and (\ref{dvu}) we
find that the abelian potential $A^\mu(x)$
in eq. (\ref{1yj1}) satisfies the Lorentz gauge condition
\bea
\partial_\mu A^\mu(x)=0.
\label{mxlz}
\eea

Similarly from eqs. (\ref{4va}), (\ref{mx7}), (\ref{dtau0}) and (\ref{dvu}) we find that
the abelian potential $A^\mu(x)$ in eq. (\ref{1yj1}) satisfies
\bea
&&~\partial^\nu \partial_\nu { A}^\mu(x)=0
\label{dda}
\eea
which gives by using eq. (\ref{mxlz})
\bea
\partial_\nu { F}^{\nu \mu}(x)=0
\label{mx8fa}
\eea
where ${ F}^{\mu \nu}(x)$ is given by eq. (\ref{fmn}).

Note that from eq. (\ref{lc}) one finds that the world line of the electron $X(\tau)$ intersects
the light cone of the observation point $x^\mu$ of the electromagnetic field only at two points, one
earlier and one later than $x_0$. From eqs. (\ref{lc}) and (\ref{rett}) [or from eq. (\ref{mx7})] one
finds that the earlier point, $X^\mu(\tau_0)$, is the only part of the path that contributes to
$A^\mu(x)$ in eq. (\ref{1yj1}) at $x^\mu$ \cite{jackson}. Since $X^\mu(\tau_0)$ is the
time-space position of the electron which produced the $A^\mu(x)$ in eq. (\ref{1yj1}) and since the
electromagnetic wave travels exactly at speed of light and the electron can not travel exactly at speed
of light (because it has finite mass even if it is very small) one finds that at the time
$t=\frac{x_0}{c}=\frac{X_0(\tau)}{c}>\frac{X_0(\tau_0)}{c}$ of $A^\mu(x)$ [see eq. (\ref{rett})]
the spatial position ${\vec x}$ of $A^\mu(x)$ is given by ${\vec x}\neq {\vec X(\tau)}$ at which
we find from eq. (\ref{mxcz}) that $j^\mu(x)=0$. Hence one finds from eq. (\ref{mx8fa}) that the
abelian potential ${ A}^\mu(x)$ in eq. (\ref{1yj1}) satisfies the Maxwell equation
\bea
\partial_\nu { F}^{\nu \mu}(x)=j^\mu(x).
\label{mxeq}
\eea

Writing four-velocity in terms of three-velocity
\bea
u^\mu = \gamma ~c~ \beta^\mu = \gamma ~c~(\beta_0, {\vec \beta})=\gamma ~c~(1, {\vec \beta}),~~~~~~~~~~~{\vec \beta}=\frac{{\vec v}}{c},~~~~~~~~~~~~\gamma = \frac{1}{\sqrt{1-{\vec \beta}^2}}
\label{ub1}
\eea
we find from eq. (\ref{1yj1}) that the Lienard-Weichert potential is given by
\bea
A^\mu(x) = e \frac{\beta^\mu(\tau_0)}{\beta(\tau_0) \cdot (x-X(\tau_0))}.
\label{uc1}
\eea
When $\beta^\mu=(\beta_0,{\vec \beta})=(1,0,0,0)$ we find from eq. (\ref{uc1})
\bea
A_0(x) =  \frac{e}{x_0-X_0}=\frac{e}{|{\vec x}-{\vec X}|}
\label{col}
\eea
which reproduces the Coulomb potential.

\subsection{ U(1) Pure Gauge Potential Produced by Electron at its Highest Speed $v\sim c$ }

As we have mentioned in the introduction, the electric charge $e$ in uniform motion
at its highest speed (which is arbitrarily close to the speed of light $v\sim c$) produces
U(1) (approximate) pure gauge potential \cite{sterman}. This can be shown from the Lienard-Weichert
potential from eq. (\ref{1yj1}) as follows.

For constant speed $v$ of the electron in uniform motion with four-velocity $u^\mu$ we find
from eq. (\ref{1yj1}) that the electromagnetic potential $A^\mu(x)$ produced by the constant
electric charge $e$ of the electron is given by
\bea
A^\mu(x) = e \frac{u^\mu}{u \cdot (x-X(\tau_0))}.
\label{5ix}
\eea
From eqs. (\ref{5ix}), (\ref{fmn}), (\ref{4va}), (\ref{dtau0}) and (\ref{dvu}) we find that the
electromagnetic field (the Maxwell field tensor) produced by the electric charge $e$ of the electron
in uniform motion with four-velocity $u^\mu$ is given by \cite{jackson}
\bea
F^{\mu \nu}(x)=ec^2\frac{(x-X(\tau_0))^\mu u^\nu -(x-X(\tau_0))^\nu u^\mu }{[u\cdot (x-X(\tau_0))]^3}.
\label{fmnu}
\eea

As mentioned in the introduction, since an electron has non-zero mass (even if negligibly small), it can not travel
exactly at speed of light $v=c$ but its highest speed can be arbitrarily close to the speed of the light $v\sim c$.
On the other hand gluon is massless and travels exactly at speed of light. Hence for a
massless particle the four velocity vector $\beta^\mu_c$ is exactly at the speed of light and it remains always at
the speed of light where $\beta^2_c=0$. Since a quark has non-zero mass (even if the light quark mass is very small),
it can not travel exactly at speed of light $v=c$ but the highest speed of the quark can be arbitrarily close to
the speed of light $v\sim c$.

For the highest speed of the electron (which is arbitrarily close to the speed of light $v \sim c$) we write
\bea
\beta^\mu_{\sim c} = (1,~{\vec \beta}_{\sim c}),~~~~~~~~~~~~~{\vec \beta}^2_{\sim c}=\frac{{\vec v}^2}{c^2}~ \sim 1,~~~~~~~~~~~~~\beta^2_{\sim c} ~\sim 0 ~~~~~~~~~~~~~~~{\rm but}~~~~~~~~~~~~~~~\beta^2_{\sim c} \neq 0.\nonumber \\
\label{bsc1}
\eea
From eq. (\ref{bsc1}) we find
\bea
u_{\sim c}^\mu = \gamma_{\sim c} \times c \times \beta^\mu_{\sim c},~~~~~~~\gamma_{\sim c} = \frac{1}{\sqrt{1-{\vec \beta}^2_{\sim c}}}\sim \infty, ~~~~~~~{\rm but}~~~~~~~\gamma_{\sim c} = \frac{1}{\sqrt{1-{\vec \beta}^2_{\sim c}}} \neq \infty.
\label{usc1a}
\eea
Note that since electron has non-zero mass we find (see eq. (\ref{usc1a}))
\bea
u^\mu_{\sim c} \neq (\infty,~\infty,~\infty,~\infty).
\label{ghf}
\eea

From eqs. (\ref{5ix}), (\ref{bsc1}), (\ref{usc1a}) and (\ref{ghf}) we find that the electron in uniform motion
at its highest speed (which is arbitrarily close to the speed of light $v \sim c$) produces electromagnetic potential
\bea
A^\mu(x) = e \frac{u^\mu_{\sim c}}{u_{\sim c} \cdot (x-X(\tau_0))}= e \frac{\beta^\mu_{\sim c}}{\beta_{\sim c} \cdot (x-X(\tau_0))}.
\label{a5ixc}
\eea
From the expression $A^\mu(x) = e \frac{\beta^\mu_{\sim c}}{\beta_{\sim c} \cdot (x-X(\tau_0))}$
from eq. (\ref{a5ixc}) we find that \cite{jackson} (see eq. (\ref{fmnu}))
\bea
F^{\mu \nu}(x)=\frac{e}{\gamma^2_{\sim c}}\frac{(x-X(\tau_0))^\mu \beta^\nu_{\sim c} -(x-X(\tau_0))^\nu \beta^\mu_{\sim c} }{[\beta_{\sim c}\cdot (x-X(\tau_0))]^3} \propto \gamma_{\sim c}
\label{fmninf}
\eea
at the spatial position ${\vec x}$ transverse to the motion of the electron (${\vec \beta}_{\sim c}\cdot {\vec x} = 0$)
at the time of closest approach (this corresponds to $t=\frac{x_0}{c}=0$ with ${\vec X}(\tau_0)={\vec \beta}_{\sim c}X_0(\tau_0)$
in eq. (\ref{fmninf})). However, at all other time-space points $x^\mu$
we find from the expression $A^\mu(x) = e \frac{\beta^\mu_{\sim c}}{\beta_{\sim c} \cdot (x-X(\tau_0))}$ from eq.
(\ref{a5ixc}) that \cite{jackson} (see eq. (\ref{fmnu}))
\bea
F^{\mu \nu}(x)=\frac{e}{\gamma^2_{\sim c}}\frac{(x-X(\tau_0))^\mu \beta^\nu_{\sim c} -(x-X(\tau_0))^\nu \beta^\mu_{\sim c} }{[\beta_{\sim c}\cdot (x-X(\tau_0))]^3} \sim 0.
\label{fmn0}
\eea
The eqs. (\ref{fmninf}) and (\ref{fmn0}) imply that at all the time-space points $x^\mu$ (except at the
spatial position ${\vec x}$ transverse to the motion of the electron at the time of closest approach) we find
(see also \cite{sterman})
\bea
A^\mu(x)= e \frac{\beta^\mu_{\sim c}}{\beta_{\sim c} \cdot (x-X(\tau_0))} \sim \partial^\mu \omega(x),~~~~~~~~~~~~~~~~\omega(x) = e ~{\rm ln}[\beta_{\sim c} \cdot (x-X(\tau_0))]
\label{pug1}
\eea
which is the expression of the abelian (approximate) pure gauge potential or U(1)
(approximate) pure gauge potential produced by the electron of constant electric
charge $e$ moving at its highest speed (which is arbitrarily close to the speed of the light $v \sim c$) where
\bea
&& \partial^\mu ~e{\rm ln}[\beta_{\sim c} \cdot (x-X(\tau_0))]=e \frac{1}{\beta_{\sim c} \cdot (x-X(\tau_0))} [\beta^\mu_{\sim c} - \beta^2_{\sim c} c \gamma_{\sim c} \partial^\mu \tau_0]\nonumber \\
&&=e \frac{1}{\beta_{\sim c} \cdot (x-X(\tau_0))} [\beta^\mu_{\sim c} - \beta^2_{\sim c} \frac{(x-X(\tau_0))^\mu}{\beta_{\sim c} \cdot (x-X(\tau_0))}]~\sim ~e \frac{\beta^\mu_{\sim c}}{\beta_{\sim c} \cdot (x-X(\tau_0))}.
\label{ue1}
\eea
We call the expression in eq. (\ref{pug1}) as the abelian (approximate) pure gauge potential or U(1)
(approximate) pure gauge potential because the electron has non-zero mass and hence it can not travel exactly at
speed of light $v=c$ or $\beta^2_c=0$ to produce the exact abelian pure gauge potential or
the exact U(1) pure gauge potential
\bea
A^\mu(x) = \partial^\mu \omega(x),~~~~~~~~~~~~{\rm where}~~~~~~~~~~~~~~\omega(x) = e ~{\rm ln}[\beta_{ c} \cdot (x-X(\tau_0))].
\label{ephg1s}
\eea

It is important to observe from eqs. (\ref{a5ixc}), (\ref{fmninf}) and (\ref{fmn0}) that the same expression
$A^\mu(x) = e \frac{\beta^\mu_{\sim c}}{\beta_{\sim c} \cdot (x-X(\tau_0))}$ in eq. (\ref{a5ixc}) which gives
$F^{\mu \nu}(x) \propto \gamma_{\sim c}$ [see eq. (\ref{fmninf})] at the spatial position ${\vec x}$ transverse to the
motion of the electron at the time of closest approach also gives $F^{\mu \nu}(x) \sim 0$ [see eq. (\ref{fmn0})]
at all other time-space points $x^\mu$ where it can be written in the form of the U(1) (approximate)
pure gauge potential as given by eq. (\ref{pug1}), see also \cite{sterman}.
Hence one finds that the exact expression of the electromagnetic potential
$A^\mu(x) = e \frac{\beta^\mu_{\sim c}}{\beta_{\sim c} \cdot (x-X(\tau_0))}$ as given by eq.  (\ref{a5ixc})
produced by the electron in uniform motion at its highest speed (which is arbitrarily close to the speed of light $v\sim c$)
at all the time-space points $x^\mu$ can be obtained from the expression of the U(1)
(approximate) pure gauge potential $A^\mu(x) = e \frac{\beta^\mu_{\sim c}}{\beta_{\sim c} \cdot (x-X(\tau_0))}$
from eq. (\ref{pug1}) even if the expression $A^\mu(x) = e \frac{\beta^\mu_{\sim c}}{\beta_{\sim c} \cdot (x-X(\tau_0))}$
can be written in the form of the U(1) (approximate) pure gauge potential
$A^\mu(x)\sim \partial^\mu \omega(x)$ at all the time-space points $x^\mu$ (except
at the spatial position ${\vec x}$ transverse to the motion of the electron at the time of closest approach).

Since the Yang-Mills theory was developed in analogy to Maxwell theory by extending U(1) group to SU(3) group
appropriately (see \cite{yang}) one expects the above features to be valid in Yang-Mills theory as well. Hence
one expects that the expression of the Yang-Mills potential $A^{\mu a}(x)$ produced by the quark in uniform
motion at its highest speed (which is arbitrarily close to the speed of light $v\sim c$) can be obtained from
the expression of the SU(3) (approximate) pure gauge potential produced by the quark.

\section{ General Expression of the SU(3) Pure Gauge Potential Produced by the Quark}

In the previous section we saw that the expression of the electromagnetic potential $A^\mu(x)$ as given by
eq. (\ref{a5ixc}) produced by the electric charge $e$ of the electron in uniform motion at its highest speed
(which is arbitrarily close to the speed of light $v\sim c$) in Maxwell theory can be obtained from
the expression of the U(1) (approximate) pure gauge potential as given by eq. (\ref{pug1}) produced by the electron.
As described in detail in the introduction, since Yang-Mills theory was developed by making analogy with Maxwell
theory by extending U(1) group to SU(3) group appropriately (see \cite{yang}), one expects by making analogy with
the Maxwell theory that the expression of the Yang-Mills potential $A^{\mu a}(x)$ produced by the color charges
$q^a(\tau)$ of the quark in uniform motion at its highest speed (which is arbitrarily close to the speed of light
$v\sim c$) in the Yang-Mills theory can be obtained from the expression of the SU(3) (approximate) pure gauge
potential produced by the quark (see section IVC for details).

Note that the U(1) pure gauge potential
$A^\mu(x)=\partial^\mu \omega(x)$ in Maxwell theory is linearly proportional
to $\omega(x)$ whereas the SU(3) pure gauge potential
$A^{\mu a} (x)=[\partial^\mu \omega^b(x)]~ [\frac{e^{gM(x)}-1}{gM(x)}]_{ab}$
in eq. (\ref{su3pure}) in Yang-Mills theory contains infinite powers of $\omega^a(x)$
where $M_{ab}(x)=f^{abc}\omega^c(x)$ is given by eq. (\ref{mab}). Note that if
we assume all the $f^{abc}=0$ then the Yang-Mills theory reduces to Maxwell-like theory.
For example, if we assume all the $f^{abc}=0$ then the SU(3) pure gauge potential
$A^{\mu a} (x)=[\partial^\mu \omega^b(x)]~ [\frac{e^{gM(x)}-1}{gM(x)}]_{ab}$
in eq. (\ref{su3pure}) becomes equal to
$\partial^\mu \omega^a(x)$. Hence in Yang-Mills theory one finds from the SU(3) pure gauge potential
$A^{\mu a} (x)=[\partial^\mu \omega^b(x)]~ [\frac{e^{gM(x)}-1}{gM(x)}]_{ab}=\partial^\mu \omega^a(x) + \frac{g}{2!} f^{abc} \omega^c(x)\partial^\mu \omega^b(x)+\frac{g^2}{3!}f^{acd}\omega^d(x)f^{cbh} \omega^h(x)  \partial^\mu \omega^b(x)+.... $~~~~ that the first term
$\partial^\mu \omega^a(x)$ corresponds to U(1) pure gauge potential $\partial^\mu \omega(x)$ in Maxwell theory.
This implies that since $\partial^\mu \omega(x)$ of the electron in eq. (\ref{pug1})
is linearly proportional to the fundamental electric charge $e$ of the electron, the
$\partial^\mu \omega^a(x)$ of the quark is linearly proportional to
the fundamental color charge $q^a(\tau)$ of the quark.
Since the $\partial^\mu \omega^a(x)$ of the quark is linearly proportional to fundamental
color charge $q^a(\tau)$ of the quark, the expression of $\omega^a(x)$ of the quark may be
calculated by using abelian approximations [see eq. (\ref{phg2}) and section IVB for details],
similar to the derivation of $\omega(x)$ of the electron in eq. (\ref{pug1}). This implies that
the general expression of the SU(3) pure gauge potential produced by the quark may be obtained
from the form of the SU(3) pure gauge potential
$A^{\mu a} (x)=[\partial^\mu \omega^b(x)]~ [\frac{e^{gM(x)}-1}{gM(x)}]_{ab}$
from eq. (\ref{su3pure}) in Yang-Mills theory by using the general expression of the
$\omega^a(x)$ of the quark by using the abelian approximations [see eq. (\ref{phg2})]
where $M_{ab}(x)=f^{abc}\omega^c(x)$ is given by eq. (\ref{mab}) [see section IVC for details].
Since the general expression of the color potential (Yang-Mills potential) produced by the
quark can be obtained from the general expression of the SU(3) pure gauge potential
produced by the quark (see section V for details) one finds that the general expression of the potential
produced by the quark may be obtained from the abelian approximations mentioned above.

Hence all it boils down is to find the general expression of the $\omega^a(x)$ of the quark
in the Yang-Mills theory. Note that, as mentioned earlier, the expression of the $\omega(x)$
of the electron in Maxwell theory is given by eq. (\ref{pug1}).

\subsection{ Abelian-Like Pure Gauge Color Potential Produced by Constant Color Charge }

However, as we will see below, it is not straightforward to extend the derivation of $\omega(x)$
in eq. (\ref{pug1}) for constant electric charge $e$ of the electron to the derivation of
$\omega^a(x)$ for time dependent color charge $q^a(\tau)$ of the quark. For this reason we will
first consider the constant color charge $q^a$ before considering the time dependent color charge
$q^a(\tau)$. Note that the constant color charge $q^a$ does not correspond to
any physical situation because the color charge $q^a(\tau)$ of the quark is time dependent. The only reason
we have considered the constant color charge $q^a$ here is to motivate similar (abelian-like)
derivation for the time dependent color charge $q^a(\tau)$ of the quark (see section IVB).

Since the electric charge $e$ is constant in Maxwell theory, we find by extending eq. (\ref{mxcz})
to constant color charge $q^a$ \cite{jackson}
\bea
{\cal J}^{\mu a}(x)=\int d\tau ~q^a~u^\mu(\tau)~\delta^{(4)}(x-X(\tau))
\label{mx4}
\eea
which is exactly similar to eq. (\ref{mxcz}) except that the constant electric
charge $e$ is replaced by constant color charge $q^a$. Since we have called the $j^\mu(x)$ produced
by the constant electric charge $e$ in eq. (\ref{mxcz}) as abelian electric current density, we will
call the ${\cal J}^{\mu a}(x)$ produced by the constant color charge $q^a$ in eq. (\ref{mx4}) as
abelian-like color current density. The ${\cal J}^{\mu a}(x)$ in eq. (\ref{mx4}) satisfies the
continuity equation
\bea
\partial_\mu {\cal J}^{\mu a}(x) =0
\label{mx5}
\eea
which confirms that the color charge $q^a$ is constant.

Using the abelian-like color current density ${\cal J}^{\mu a}(x)$ from eq. (\ref{mx4}) in the
inhomogeneous wave equation
\bea
\partial^\nu \partial_\nu {\cal A}^{\mu a}(x)={\cal J}^{\mu a}(x)
\label{mx2v}
\eea
we find the solution \cite{jackson}
\bea
{\cal A}^{\mu a}(x)=q^a\frac{u^\mu(\tau_0)}{u(\tau_0) \cdot (x-X(\tau_0))}
\label{mx6}
\eea
by using ${\cal J}^{\mu a}(x)$ from eq. (\ref{mx4}) in
\bea
{\cal A}^{\mu a}(x)=\int d^4x' D_r(x-x'){\cal J}^{\mu a}(x')
\eea
where $D_r(x-x')$ is the retarded greens function.

Note that we have used the curly notations ${\cal J}^{\mu a}(x),~{\cal A}^{\mu a}(x)$ etc. in
the abelian-like case for a constant color charge $q^a$ in this section and in the abelian-like
non-abelian case for a time dependent color charge $q^a(\tau)$ in section IVB but we have used
the usual notations $j^{\mu a}(x),~A^{\mu a}(x)$ etc. in the (full) non-aelian case in Yang-Mills
theory in sections I, IVC, V and in appendix C. The eq. (\ref{mx2v}) is similar to eq. (\ref{mxpz})
in Maxwell theory and the eq. (\ref{mx6}) is similar to eq. (\ref{1yj1}) in Maxwell theory.

From eq. (\ref{mx8a}) in the appendix we find that ${\cal A}^{\mu a}(x)$ as given by eq.
(\ref{mx6}) satisfies the Lorentz gauge condition
\bea
\partial_\mu {\cal A}^{\mu a}(x)=0.
\label{mx8b}
\eea
From eq. (\ref{mx8d}) in the appendix we find that ${\cal A}^{\mu a}(x)$ as given by eq.
(\ref{mx6}) satisfies the equation
\bea
\partial^\nu \partial_\nu {\cal A}^{\mu a}(x) = 0.
\label{mx8e}
\eea
From eqs. (\ref{mx8e}) and (\ref{mx8b}) we find
\bea
\partial_\nu {\cal F}^{\nu \mu a}(x)=0
\label{mx8f}
\eea
where
\bea
{\cal F}^{\mu \nu a}(x) =\partial^\mu {\cal A}^{\nu a}(x) -\partial^\nu {\cal A}^{\mu a}(x).
\label{afmn}
\eea
For a fermion of non-zero mass having color charge $q^a$ one finds [see the discussion above eq. (\ref{mxeq})]
that at the time $t=\frac{x_0}{c}=\frac{X_0(\tau)}{c}>\frac{X_0(\tau_0)}{c}$ of ${\cal A}^{\mu a}(x)$ [see eq.
(\ref{rett})] the spatial position ${\vec x}$ of ${\cal A}^{\mu a}(x)$ in eq. (\ref{mx6}) is given by
${\vec x}\neq {\vec X}(\tau)$ at which one finds from eq. (\ref{mx4}) the ${\cal J}^{\mu a}(x)=0$ which implies
that the eq. (\ref{mx8f}) satisfies the Maxwell-like equation
\bea
\partial_\nu {\cal F}^{\nu \mu a}(x)={\cal J}^{\mu a}(x).
\label{mx1}
\eea

Note that the expression of ${\cal A}^{\mu a}(x)$ in eq. (\ref{mx6}) is exactly same as the abelian
potential $A^\mu(x)$ produced by the constant electric charge $e$ in eq. (\ref{1yj1}) except
that the constant electric charge $e$ is replaced by the constant color charge $q^a$. Furthermore
the ${\cal A}^{\mu a}(x)$ in eq. (\ref{mx6}) satisfies Lorenz gauge condition as given by eq. (\ref{mx8b})
which is similar to the Lorenz gauge condition in abelian theory as given by eq. (\ref{mxlz}).
In addition to this the ${\cal A}^{\mu a}(x)$ in eq. (\ref{mx6}) satisfies the Maxwell-like equation
as given by eq. (\ref{mx1}) which is similar to Maxwell equation as given by eq. (\ref{mxeq})
in abelian theory. For these reasons, since we have called $A^\mu(x)$ in eq. (\ref{1yj1}) as the abelian potential
produced by the constant electric charge $e$, we will call ${\cal A}^{\mu a}(x)$ in eq. (\ref{mx6}) as the
abelian-like color potential produced by the constant color charge $q^a$.

When the speed of the color charge is arbitrarily close to the speed of light ($v \sim c$) we find
from eqs. (\ref{mx6}) and (\ref{afmn}) [by using eqs. (\ref{4va}), (\ref{dtau0}) and (\ref{dvu})]
that at all the time-space points $x^\mu$ (except at the
spatial position ${\vec x}$ transverse to the motion of the color charge at the time of closest approach)
the
\bea
{\cal F}^{\mu \nu a}(x) \sim 0
\label{afmn0}
\eea
which is similar to eq. (\ref{fmn0}) in abelian theory.
From eq. (\ref{afmn0}) we find (similar to eq. (\ref{pug1}) in abelian theory) that the abelian-like color
potential ${\cal A}^{\mu a}(x)$ in eq. (\ref{mx6}) at all the time-space points $x^\mu$ (except at the
spatial position ${\vec x}$ transverse to the motion of the color charge at the time of closest approach)
produced by the constant color charge $q^a$ at its highest speed (which is arbitrarily close to the speed
of light $v \sim c$) can be written in the form
\bea
{\cal A}^{\mu a}(x)= q^a \frac{\beta^\mu_{\sim c}}{\beta_{\sim c} \cdot (x-X(\tau_0))} \sim \partial^\mu \omega^a(x),~~~~~~~{\rm where}~~~~~~~~~~~~\omega^a(x)= q^a~ {\rm ln}[\beta_{\sim c} \cdot (x-X(\tau_0))]. \nonumber \\
\label{phn1}
\eea
Note that the expression of ${\cal A}^{\mu a}(x)$ in eq. (\ref{phn1}) is exactly same as the expression of
$A^\mu(x)$ in eq. (\ref{pug1}) except that the constant electric charge $e$ is replaced by the constant color
charge $q^a$. Since we have called $A^\mu(x)$ in eq. (\ref{pug1}) as the abelian (approximate) pure gauge potential
produced by the constant electric charge $e$ in uniform motion at its highest speed (which is arbitrarily close
to the speed of light $v\sim c$), we will call ${\cal A}^{\mu a}(x)$ in eq. (\ref{phn1}) as the
abelian-like (approximate) pure gauge color potential produced by the constant color charge $q^a$
in uniform motion at its highest speed (which is arbitrarily close to the speed of light $v\sim c$).

By using $\omega^a(x)$ from eq. (\ref{phn1}) in eq. (\ref{su3pure}) we find that the non-abelian
SU(3) pure gauge potential
\bea
A^{\mu a} (x)=[\partial^\mu \omega^b(x)]~ [\frac{e^{gM(x)}-1}{gM(x)}]_{ab}=[\partial_\mu \omega^b(x)]~ [1 + \frac{g}{2!}M(x)
+\frac{g^2}{3!}M^2(x)+\frac{g^3}{4!} M^3(x)  +...]_{ab}=\partial_\mu \omega^a(x)\nonumber \\
\label{su3purev}
\eea
becomes an abelian-like pure gauge color potential $\partial^\mu \omega^a(x)$ where $M_{ab}(x)$ is given by eq. (\ref{mab}). Note that the
repeated color indices $b$(=1,2,....8) are summed in eq. (\ref{su3purev}).

Hence we find from eq. (\ref{su3purev}) that if color charge $q^a$ is constant then the SU(3) pure gauge
potential $A^{\mu a}(x)$ obtained from the SU(3) pure gauge equation
$T^aA^{\mu a}(x)=\frac{1}{ig}[\partial^\mu U(x)]U^{-1}(x)$ reduces to an abelian-like pure gauge color potential
$A^{\mu a}(x)=\partial^\mu \omega^a(x)$ where $U(x)=e^{igT^a\omega^a(x)}$.

This is expected because when the color charge $q^a$ is constant, the entire physics
becomes similar to QED except that the role of constant electric charge $e$ is replaced by constant
color charge $q^a$, where $a$=1,2,...8. This is equivalent to an abelian-like continuity equation
as given by eq. (\ref{mx5}) which becomes possible only when all the generators $T^a$ are diagonal
{\it i.e.}, when $f^{abc}=0$ for all values of $a,b,c$=1,2,...8.

\subsection{ Abelian-Like Non-Abelian Pure Gauge Color Potential Produced by Time Dependent Color Charge }

However, since all the the generators $T^a$ in SU(3) group are not diagonal
the non-abelian Yang-Mills color current density $j^{\mu a}(x)$ of the quark does not satisfy the continuity
equation similar to that in eq. (\ref{mx5}). Instead, the non-abelian Yang-Mills color current density
$j^{\mu a}(x)$ of the quark satisfies the eq. (\ref{ab}) which implies that the color charge of the quark is
not constant.

As discussed in the introduction since the color charge of the quark is not constant and it is
independent of space coordinate ${\vec x}$ one finds that the color charge of the quark in Yang-Mills
theory has to be time dependent.

Hence we find that unlike constant color charge $q^a$ which does not satisfy eq. (\ref{ab}), the
time dependent color charge $q^a(t)$ satisfies eq. (\ref{ab}).

As mentioned earlier the constant color charge $q^a$ does not correspond to any physical situation
and we have considered it here only to motivate similar (abelian-like) derivation for the time dependent
color charge $q^a(\tau)$. Hence in order to derive the general expression of $\omega^a(x)$ of the
quark in terms of the time dependent color charge $q^a(\tau)$ of the quark we proceed as follows.
Extending eq. (\ref{mx4}) to time dependent color charge $q^a(\tau)$ we write
\bea
{\cal J}^{\mu a}(x)=\int d\tau ~q^a(\tau)~u^\mu(\tau)~\delta^{(4)}(x-X(\tau))
\label{mx8}
\eea
which gives
\bea
\partial_\mu {\cal J}^{\mu a}(x)=\int d\tau ~\frac{dq^a(\tau)}{d\tau}~\delta^{(4)}(x-X(\tau))
\label{mx8fmn}
\eea
which is non-zero at $x^\mu =X^\mu(\tau)$ which confirms that the color charge $q^a(\tau)$ is not constant.
Using ${\cal J}^{\mu a}(x)$ from eq. (\ref{mx8}) in the inhomogeneous wave equation
\bea
\partial^\nu \partial_\nu {\cal A}^{\mu a}(x)={\cal J}^{\mu a}(x)
\label{nabp}
\eea
we find the solution \cite{jackson}
\bea
{\cal A}^{\mu a}(x)=q^a(\tau_0)\frac{u^\mu(\tau_0)}{u(\tau_0) \cdot (x-X(\tau_0))}
\label{mx10}
\eea
by using ${\cal J}^{\mu a}(x)$ from eq. (\ref{mx8}) in
\bea
{\cal A}^{\mu a}(x)=\int d^4x' D_r(x-x'){\cal J}^{\mu a}(x')
\label{mx3t}
\eea
where $D_r(x-x')$ is the retarded greens function.

Note that eq. (\ref{nabp}) is similar to eq. (\ref{mx2v}) for the constant color charge $q^a$ case
except that constant color charge $q^a$ is replaced by time dependent color charge $q^a(\tau)$. Hence
one finds that eq. (\ref{mx10}) is similar to eq. (\ref{mx6}) for the constant color charge $q^a$ case
except that constant color charge $q^a$ is replaced by time dependent color charge $q^a(\tau_0)$.
However, the ${\cal A}^{\mu a}(x)$ in eq. (\ref{mx6}) for the constant color charge $q^a$ case satisfies
the Lorentz gauge condition as given by eq. (\ref{mx8b}) whereas the ${\cal A}^{\mu a}(x)$ in eq. (\ref{mx10})
for the time dependent color charge $q^a(\tau)$ case does not satisfy the Lorentz gauge condition.
Hence one finds that the ${\cal A}^{\mu a}(x)$ in eq. (\ref{mx10})
for the time dependent color charge $q^a(\tau)$ case does not satisfy the Maxwell-like equation (similar to
eq. (\ref{mx1}) for the constant color charge $q^a$ case).
This is the main difference between constant color charge $q^a$ case and time dependent color charge
$q^a(\tau)$ case. However, as we will see below, when the time dependent
color charge $q^a(\tau)$ in uniform motion is at its highest speed (which is arbitrarily close to the
speed of light $v\sim c$) the ${\cal A}^{\mu a}(x)$ in eq. (\ref{mx10}) satisfies Lorentz gauge
condition and hence satisfies Maxwell-like equation (similar to eq. (\ref{mx1}) for constant
color charge case) at all the time-space points $x^\mu$ (except at the spatial position ${\vec x}$
transverse to the motion of the color charge at the closest time of approach) with the ${\cal F}^{\mu \nu a}(x)$
given by
\bea
{\cal F}^{\mu \nu a}(x) =\partial^\mu {\cal A}^{\nu a}(x) -\partial^\nu {\cal A}^{\mu a}(x).
\label{afmnt}
\eea
Hence when the time dependent color charge in uniform motion
is at its highest speed (which is arbitrarily close to the speed of light $v\sim c$)
one expects to derive the general expression of $\omega^a(x)$ of the
time dependent color charge $q^a(\tau)$ at all the time-space points $x^\mu$
(except at the spatial position ${\vec x}$
transverse to the motion of the color charge at the closest time of approach)
[similar to the derivation of the $\omega^a(x)$ in eq. (\ref{phn1})
for the constant color charge $q^a$ case]. This can be shown as follows.

From eq. (\ref{mxd}) in the appendix we find that ${\cal A}^{\mu a}(x)$ in eq. (\ref{mx10}) satisfies
\bea
\partial^\nu \partial_\nu {\cal A}^{\mu a}(x) =0.
\label{mxe}
\eea
Similar to the electromagnetic case [see the paragraph above eq. (\ref{mxeq})],
since $X^\mu(\tau_0)$ is the time-space position of the color charge of the quark
which produced ${\cal A}^{\mu a}(x)$ in eq. (\ref{mx10}) and since a quark
(even if the mass of the light quark is very small) can not travel exactly at speed of light
one finds that at the time $t=\frac{x_0}{c}=\frac{X_0(\tau)}{c}>\frac{X_0(\tau_0)}{c}$
of ${\cal A}^{\mu a}(x)$ [see eq. (\ref{rett})] the spatial position ${\vec x}$ of ${\cal A}^{\mu a}(x)$ is
given by ${\vec x}\neq {\vec X(\tau)}$ at which we find from eqs. (\ref{mx10}), (\ref{afmnt}) and (\ref{mx8fmn}) that
\bea
\partial_\mu \partial_\nu {\cal F}^{\mu \nu a}(x)=0=\partial_\mu {\cal J}^{\mu a}(x).
\label{mxfmn}
\eea
One should not be confused with the fact that $\partial_\mu {\cal J}^{\mu a}(x)$ can be
non-zero in eq. (\ref{mx8fmn}) whereas $\partial_\mu {\cal J}^{\mu a}(x)= 0$ in eq. (\ref{mxfmn}). This is because the time-space
point $x^\mu$ at which  $\partial_\mu {\cal J}^{\mu a}(x)\neq 0$ in eq. (\ref{mx8fmn}) is given by
$x^\mu =X^{\mu}(\tau)$
whereas the spatial position ${\vec x}$ at which $\partial_\mu {\cal J}^{\mu a}(x)= 0$ in eq.
(\ref{mxfmn}) is given by ${\vec x}\neq {\vec X}(\tau)$ [see the discussions above eq. (\ref{mxfmn})].
Similarly one finds that the $\partial_\mu {\cal J}^{\mu a}(x)= 0$
in eq. (\ref{mxfmn}) does not mean that the color charge is constant because
$\partial_\mu {\cal J}^{\mu a}(x)= 0$ in eq. (\ref{mxfmn}) is at the spatial position ${\vec x}\neq {\vec X}(\tau)$
whereas at the time-space position $x^\mu = X^\mu(\tau)$ one finds from eq. (\ref{mx8fmn}) that
$\partial_\mu {\cal J}^{\mu a}(x)\neq 0$ which confirms that the
color charge $q^a(\tau)$ is time dependent.

From eq. (\ref{mxa}) in the appendix we find that ${\cal A}^{\mu a}(x)$ in eq. (\ref{mx10}) gives
\bea
\partial_\mu {\cal A}^{\mu a}(x) = \frac{{\dot q}^a(\tau_0)}{u(\tau_0) \cdot (x-X(\tau_0))}
\label{mxhi}
\eea
which does not satisfy the Lorentz gauge condition where
\bea
{\dot q}^a(\tau_0) = \frac{dq^a(\tau)}{d\tau}|_{\tau=\tau_0}.
\label{qqpi}
\eea

Note that eq. (\ref{mxe}) is consistent with eq. (\ref{nabp})
because the spatial position ${\vec x}$ of ${\cal A}^{\mu a}(x)$ in eq. (\ref{mx10}) is
given by ${\vec x}\neq {\vec X(\tau)}$ [see the discussion above eq. (\ref{mxfmn})] where
one finds from eq. (\ref{mx8}) the ${\cal J}^{\mu a}(x)=0$. However, since Lorentz gauge
condition is not satisfied in eq. (\ref{mxhi}) one finds that the ${\cal A}^{\mu a}(x)$
in eq. (\ref{mx10}) does not satisfy the Maxwell-like equation
(similar to eq. (\ref{mx1}) for the constant color charge $q^a$ case) where $ {\cal F}^{\mu \nu a}(x)$ is
given by eq. (\ref{afmnt}). For unform velocity we find from eq. (\ref{mxhi})
\bea
\partial_\mu {\cal A}^{\mu a}(x) = \frac{{\dot q}^a(\tau_0)}{u \cdot (x-X(\tau_0))}.
\label{mxh}
\eea
When the time dependent color charge $q^a(\tau)$ in uniform motion is at its highest
speed (which is arbitrarily close to the speed of light $v \sim c$) we find
from eq. (\ref{mx10}) that the ${\cal A}^{\mu a}(x)$ is given by
\bea
{\cal A}^{\mu a}(x)=q^a(\tau_0)\frac{\beta^\mu_{\sim c}}{\beta_{\sim c} \cdot (x-X(\tau_0))}.
\label{mxk}
\eea
Hence we find from eq. (\ref{mxh}) that at all the time-space points $x^\mu$
(except at the spatial position ${\vec x}$ transverse to the motion of the color charge at the
time of closest approach) the ${\cal A}^{\mu a}(x)$ in eq. (\ref{mxk}) satisfies (approximate) Lorentz
gauge condition
\bea
\partial_\mu {\cal A}^{\mu a}(x) \sim 0.
\label{mxi}
\eea
From eqs. (\ref{mxi}) and (\ref{mxe}) we find that at all the time-space
points $x^\mu$ (except at the spatial position ${\vec x}$ transverse to the motion of the color charge
at the time of closest approach) the ${\cal A}^{\mu a}(x)$ in eq. (\ref{mxk}) satisfies the equation
\bea
\partial_\nu {\cal F}^{\nu \mu a}(x)\sim 0
\label{mxka}
\eea
where ${\cal F}^{\mu \nu a}(x)$ is given by eq. (\ref{afmnt}). Since the spatial position
${\vec x}$ of ${\cal A}^{\mu a}(x)$ is given by ${\vec x}\neq {\vec X(\tau)}$
[see the discussion above eq. (\ref{mxfmn})] at which the ${\cal J}^{\mu a}(x)=0$ [see eq.
(\ref{mx8})] we find that the eq. (\ref{mxka}) satisfies Maxwell-like equation
$\partial_\nu {\cal F}^{\nu \mu a}(x)={\cal J}^{\mu a}(x)$ at all the time-space points $x^\mu$
(except at the spatial position ${\vec x}$ transverse to the motion of the color charge at the time of closest approach).

From eq. (\ref{mxp}) in the appendix we find at all the
time-space points $x^\mu$ (except at the spatial position ${\vec x}$ transverse to the motion of the
color charge at the time of closest approach) that the ${\cal A}^{\mu a}(x)$ in eq. (\ref{mxk}) gives
\bea
{\cal F}^{\mu \nu a}(x)\sim 0
\label{mxq}
\eea
where ${\cal F}^{\mu \nu a}(x)$ is given by eq. (\ref{afmnt}). Eq. (\ref{mxq}) for time dependent color
charge $q^a(\tau)$ case is similar to eq. (\ref{afmn0}) for the constant color charge $q^a$ case.

We find from eqs. (\ref{mxk}), (\ref{mxi}), (\ref{mxka}) and (\ref{mxq})
that at all the time-space points $x^\mu$ (except at the spatial position ${\vec x}$
transverse to the motion of the color charge at the time of closest approach)
the ${\cal A}^{\mu a}(x)$ in eq. (\ref{mxk}) produced by the time dependent color charge $q^a(\tau)$
in uniform motion at its highest speed (which is arbitrarily close to the speed of light $v\sim c$) satisfies
\bea
\partial_\mu {\cal A}^{\mu a}(x) \sim 0,
\label{mxt}
\eea
\bea
\partial_\mu {\cal F}^{\mu \nu a}(x)\sim 0
\label{mxu}
\eea
and
\bea
{\cal F}^{\mu \nu a}(x)\sim 0
\label{mxv}
\eea
where ${\cal F}^{\mu \nu a}(x)$ is given by eq. (\ref{afmnt}) which implies that the ${\cal A}^{\mu a}(x)$
in eq. (\ref{mxk}) can be written in the form
\bea
{\cal A}^{\mu a}(x)=q^a(\tau_0) \frac{\beta^\mu_{\sim c}}{\beta_{\sim c} \cdot (x-X(\tau_0))} \sim \partial^\mu \omega^a(x),~~~~~~~\omega^a(x) = \int dl_c~ \frac{q^a(\tau_0)}{l_c},~~~~~~~l_c=\beta_{\sim c} \cdot (x-X(\tau_0)) \nonumber \\
\label{phg2}
\eea
which is similar to eq. (\ref{phn1}) for the constant color charge $q^a$ case. Note that we have called the
expression of ${\cal A}^{\mu a}(x)$ in eq. (\ref{phn1}) as the abelian-like (approximate) pure gauge color potential
which is produced by the constant color charge $q^a$ in uniform motion at its highest speed
(which is arbitrarily close to the speed of light $v \sim c$). Similarly we will call the expression of
${\cal A}^{\mu a}(x)$ in eq. (\ref{phg2}) as the abelian-like non-abelian
(approximate) pure gauge color potential which is produced by the time dependent color charge $q^a(\tau)$
of the quark in uniform motion at its highest speed (which is arbitrarily close to the speed of light $v \sim c$).

One can view the abelian (approximate) pure gauge potential in eq. (\ref{pug1})
produced by the constant electric charge $e$ of the electron is like soft/collinear photon
field in QED which interacts with electron but not with photon. Similarly one
finds that the abelian-like (approximate) pure gauge color potential in eq. (\ref{phn1})
produced by constant color charge $q^a$ can interact with quark but not with gluon even
if it carries color indices. On the other hand the abelian-like non-abelian (approximate) pure
gauge color potential in eq. (\ref{phg2}) which is produced by the time dependent color charge
$q^a(\tau)$ of the quark is like soft/collinear gluon field in QCD which
interacts with gluon as well as with quark.

\subsection{ General Expression of the SU(3) Pure Gauge Potential Produced by the Quark }

The expression of the $\omega(x)$ of the electron in uniform motion at its highest speed (which is arbitrarily
close to the speed of light $v \sim c$) at all the time-space points $x^\mu$
(except at the spatial position ${\vec x}$ transverse to the motion of the electron at the time of closest approach)
is given by eq. (\ref{pug1}). Similarly, the general
expression of the $\omega^a(x)$ of the quark in uniform motion at its highest speed (which is arbitrarily close to the
speed of light $v \sim c$) at all the time-space points $x^\mu$ (except at the
spatial position ${\vec x}$ transverse to the motion of the quark at the time of closest approach)
is given by eq. (\ref{phg2}). As described in detail in the second paragraph
of section IV, by comparing eqs. (\ref{pug1}) and (\ref{phg2}) with eqs. (\ref{purea}) and (\ref{su3pure}) [where $M_{ab}(x)$
is given by eq. (\ref{mab})] one finds that the general expression of the SU(3) (approximate) pure gauge potential produced
by the time dependent color charges $q^a(\tau)$ of the quark in uniform motion at its highest speed (which is arbitrarily close
to the speed of light $v \sim c$) at all the time-space points $x^\mu$ (except at the
spatial position ${\vec x}$ transverse to the motion of the quark at the time of closest approach)
can be obtained from the form of the SU(3) pure gauge potential from
eq. (\ref{su3pure}) by using the general expression of the $\omega^a(x)$ of the quark [as given by eq. (\ref{phg2})] in eqs.
(\ref{mab}) and (\ref{su3pure}).

Hence by using the general expression of the $\omega^a(x)$ of the quark [as given by eq. (\ref{phg2})] in
the form of the SU(3) pure gauge potential in eq. (\ref{su3pure}) [where $M_{ab}(x)$ is given by eq. (\ref{mab})]
we find that the general expression of the SU(3) (approximate) pure gauge potential
produced by the time dependent color charges $q^a(\tau)$ of the quark in uniform motion at its highest speed (which
is arbitrarily close to the speed of light $v \sim c$) at all the time-space points $x^\mu$
(except at the spatial position ${\vec x}$ transverse to the motion of the quark at the time of closest approach)
is given by
\bea
A^{\mu a}(x) =  \frac{\beta^\mu_{\sim c}}{\beta_{\sim c} \cdot (x-X(\tau_0))}q^b(\tau_0) [\frac{e^{g\int dl_c \frac{Q(\tau_0)}{l_c}}-1}{g\int dl_c \frac{Q(\tau_0)}{l_c}}]_{ab}~ \sim [\partial^\mu \omega^b(x)]~ [\frac{e^{gM(x)}-1}{gM(x)}]_{ab} \nonumber \\
\label{spg1}
\eea
where
\bea
Q^{ab}(\tau_0) =f^{abd}q^d(\tau_0),~~~~l_c= \beta_{\sim c} \cdot (x-X(\tau_0)),~~~~M_{ab}(x)=f^{abd}\omega^d(x),~~~~\omega^a(x) = \int dl_c~ \frac{q^a(\tau_0)}{l_c}.\nonumber \\
\eea

Eq. (\ref{spg1}) in Yang-Mills theory is analogous to eq. (\ref{pug1}) in Maxwell theory.

From eqs. (\ref{spg1}), (\ref{4va}), (\ref{dtau0}) and (\ref{dvu}) we find that at all the
time-space points $x^\mu$ (except at the spatial position ${\vec x}$ transverse to the motion
of the quark at the time of closest approach) the quark in uniform motion at its
highest speed (which is arbitrarily close to the speed of light $v \sim c$) produces
\bea
F^{\mu \nu a}(x)\sim 0
\label{fpg1}
\eea
where $F^{\mu \nu a}(x)$ is given by eq. (\ref{fmna}).

Eq. (\ref{fpg1}) in Yang-Mills theory is analogous to eq. (\ref{fmn0}) in Maxwell theory.

As mentioned earlier, it is important to observe that in Maxwell theory the expression
$A^\mu(x) = e \frac{\beta^\mu_{\sim c}}{\beta_{\sim c} \cdot (x-X(\tau_0))}$
in eq. (\ref{pug1}) and the expression $A^\mu(x) = e \frac{\beta^\mu_{\sim c}}{\beta_{\sim c} \cdot (x-X(\tau_0))}$
in eq. (\ref{a5ixc}) are same. As described in detail in the introduction, the Yang-Mills theory was developed by
making analogy with the Maxwell theory by extending U(1) group to SU(3) group appropriately, see \cite{yang}.
Since the eq. (\ref{spg1}) in Yang-Mills theory is analogous to eq. (\ref{pug1}) in  Maxwell theory,
by making analogy with Maxwell theory we find from eq. (\ref{spg1}) that the expression
\bea
A^{\mu a}(x) =  \frac{\beta^\mu_{\sim c}}{\beta_{\sim c} \cdot (x-X(\tau_0))}q^b(\tau_0) [\frac{e^{g\int dl_c \frac{Q(\tau_0)}{l_c}}-1}{g\int dl_c \frac{Q(\tau_0)}{l_c}}]_{ab},~~~~~Q^{ab}(\tau_0) =f^{abd}q^d(\tau_0),~~~l_c= \beta_{\sim c} \cdot (x-X(\tau_0))\nonumber \\
\label{afnab}
\eea
in Yang-Mills theory is analogous to eq. (\ref{a5ixc}) in Maxwell theory.

From eqs. (\ref{afnab}), (\ref{fmna}), (\ref{4va}), (\ref{dtau0}) and (\ref{dvu}) we find
\bea
F^{\mu \nu a}(x) \propto \gamma_{\sim c}
\label{fpg1inf}
\eea
at the spatial position ${\vec x}$ transverse to the motion of the quark at the time of closest approach.

The eq. (\ref{fpg1inf}) in Yang-Mills theory is analogous to eq. (\ref{fmninf}) in Maxwell theory.

Note that, as mentioned earlier, the expression
$A^{\mu}(x) = e \frac{\beta^\mu_{\sim c}}{\beta_{\sim c} \cdot (x-X(\tau_0))}$ in eq. (\ref{a5ixc})
in Maxwell theory is the exact expression of the Maxwell potential (electromagnetic potential)
$A^{\mu }(x)$ at all the time-space points $x^\mu$ produced by the electric charge $e$ of the electron
in uniform motion at its highest speed (which is arbitrarily close to the speed of light $v \sim c$).
Hence by making analogy with Maxwell theory we have found that the expression $A^{\mu a}(x)=  \frac{\beta^\mu_{\sim c}}{\beta_{\sim c} \cdot (x-X(\tau_0))}q^b(\tau_0) [\frac{e^{g\int dl_c \frac{Q(\tau_0)}{l_c}}-1}{g\int dl_c \frac{Q(\tau_0)}{l_c}}]_{ab}$ in eq. (\ref{afnab})
in Yang-Mills theory is the general expression of the Yang-Mills potential $A^{\mu a}(x)$ at all the time-space points $x^\mu$
produced by the color charges $q^a(\tau)$ of the quark in uniform motion at its highest speed (which is arbitrarily close to the
speed of light $v \sim c$).

\section{ General Form of Color Potential Produced by Color Charges of the Quark }

In Maxwell theory we find from eq. (\ref{col}) that the Coulomb potential is given by
\bea
A_0(x) = e \frac{\beta_0}{\beta_0(x_0-X_0)},~~~~~~~~~~~~~~~~~~~~~~~~x_0-X_0=|{\vec x}-{\vec X}|,
\label{colf}
\eea
and from eq. (\ref{uc1}) we find that the Lienard-Weichert potential is given by
\bea
A^\mu(x) = e \frac{\beta^\mu(\tau_0)}{\beta(\tau_0) \cdot (x-X(\tau_0))},~~~~~~~~~~~~~~~x_0-X_0(\tau_0)=|{\vec x}-{\vec X}(\tau_0)|,
\label{lw}
\eea
and from eq. (\ref{pug1}) we find that the U(1) (approximate) pure gauge potential  is given by
\bea
A^\mu(x)= e \frac{\beta^\mu_{\sim c}}{\beta_{\sim c} \cdot (x-X(\tau_0))},~~~~~~~~~~~~~~~~~~~x_0-X_0(\tau_0)=|{\vec x}-{\vec X}(\tau_0)|,
\label{pglw}
\eea
where $\beta_0$ is the zeroth component of the four velocity vector,
$\beta^\mu_{\sim c}$ is the four-velocity arbitrarily close to the speed of light
($v \sim c$) and $\beta^\mu(\tau)$ is any arbitrary four velocity of the electron.

Note that as we saw in section IIIA the expression $A^\mu(x)= e \frac{\beta^\mu_{\sim c}}{\beta_{\sim c} \cdot (x-X(\tau_0))}$
in eq. (\ref{pglw}) which can be written in the form of the U(1) (approximate) pure gauge potential as given by eq. (\ref{pug1})
at all the time-space points $x^\mu$ (except at the spatial position ${\vec x}$
transverse to the motion of the electron at the time of closest approach)
is the same expression $A^\mu(x)= e \frac{\beta^\mu_{\sim c}}{\beta_{\sim c} \cdot (x-X(\tau_0))}$
in eq. (\ref{a5ixc}) produced by the electron in uniform motion
at its highest speed (which is arbitrarily close to the speed of light $v\sim c$) at all the time-space points $x^\mu$.

Hence eqs. (\ref{colf}), (\ref{lw}) and (\ref{pglw}) suggest that in Maxwell
theory (or in U(1) gauge theory) there is just one formula for the electromagnetic potential in the
nature. Depending on three different speed limits
we call them three different potentials. For example, when speed is zero $\left(\beta^\mu =(\beta_0,~0,~0,~0)\right)$
we call it Coulomb potential (see eq. (\ref{colf})), when speed is finite $\left(\beta^\mu(\tau) = (\beta_0,~{\vec \beta}(\tau))\right)$
we call it Lienard-Weichert potential (see eq. (\ref{lw})) and when
speed is arbitrarily close to the speed of light $\left(\beta^\mu =(\beta_0,~{\vec \beta}_{\sim c})\right)$
we call it U(1) (approximate) pure gauge potential (see eq. (\ref{pglw})).

Eqs. (\ref{colf}), (\ref{lw}) and (\ref{pglw}) suggest that if we know one formula then we can get the other
two formulas by changing the four-velocity vector. For example, the static Coulomb potential $A_0(x)$ in eq.
(\ref{colf}) can be obtained from the Lienard-Weichert potential $A^\mu(x)$ by replacing
\bea
\beta^\mu(\tau) \rightarrow (\beta_0,0,0,0),~~~~~~~~~~~~~{\rm or ~by ~putting}~~~~~~~~~~~~~~~~{\vec \beta}=0
\label{mya1}
\eea
in eq. (\ref{lw}). This is obvious because when we put velocity ${\vec \beta}=0$ then the particle comes to rest,
see eq. (\ref{ub1}). Similarly the static Coulomb potential $A_0(x)$ in eq. (\ref{colf})
can be obtained from the U(1) (approximate) pure gauge potential $A^\mu(x)$ by replacing
\bea
\beta^\mu_{\sim c} \rightarrow (\beta_0,0,0,0)
\label{my2}
\eea
in eq. (\ref{pglw}). Note that $\beta_0=1$ remains unchanged in $\beta^\mu(\tau)$
and in $\beta^\mu_{\sim c}$, see eqs. (\ref{ub1}) and (\ref{bsc1}). Since Yang-Mills theory is
obtained from Maxwell theory by extending the U(1) gauge group to SU(3) gauge group (see \cite{yang}),
one expects eqs. (\ref{mya1}) and (\ref{my2}) to be valid in Yang-Mills theory. Hence one
expects that the exact expression of the Yang-Mills potential $A^{\mu a}(x)$ produced by the
quark moving at any speed can be obtained from the expression of the SU(3) (approximate) pure gauge potential
produced by the quark in uniform motion at its highest speed (which is arbitrarily close to the speed of
light $v \sim c$).

In other words, one can obtain the exact expression of the Yang-Mills potential (color potential)
$A_0^a(x)$ produced in a frame where the quark is at rest from the expression of the
SU(3) (approximate) pure gauge potential $A^{\mu a}(x)$ produced in a frame where the quark
is in uniform motion at its highest speed (which is arbitrarily close to the speed of light
$v \sim c$) by using eq. (\ref{my2}) in (\ref{spg1}).

The above argument can be generalized to a frame where the charge is moving at any
arbitrary speed. For example, by comparing eq. (\ref{pglw}) with eq. (\ref{lw}) we find that the
exact expression of the physical potential (Lienard-Weichert potential) in eq. (\ref{lw}) in Maxwell
theory can be obtained from the expression of the U(1) (approximate) pure gauge potential by replacing
\bea
\beta^\mu_{\sim c} \rightarrow \beta^\mu(\tau)
\label{rep1}
\eea
in eq. (\ref{pglw}). This is expected because (approximate) pure gauge potential is obtained from the
Lienard-Wiechert potential when the speed of the electron in uniform motion approaches its highest speed
(which is arbitrarily close to the speed of light $v \sim c$), see eqs. (\ref{fmn0}) and (\ref{pug1}).
By the word "physical potential" we mean the potential which gives physical (gauge invariant) electromagnetic
field $F_{\mu \nu}$ in Maxwell theory (see eq. (\ref{fmn})) and physical (gauge invariant)
$\sum_{a=1}^8 F_{\mu \nu}^a F^{\mu \nu a}$ (see eq. (\ref{fmna})) in Yang-Mills theory.

The replacement as given by eq. (\ref{my2})
can also be explicitly verified by using Lorentz transformations.

As mentioned earlier (see section IIIA) the expression
$A^\mu(x)= e \frac{\beta^\mu_{\sim c}}{\beta_{\sim c} \cdot (x-X(\tau_0))}$ in
eq. (\ref{a5ixc}) which is valid at all the time-space points $x^\mu$
can be written in the form of the U(1) (approximate) pure gauge potential
at all the time-space points
$x^\mu$ (except at the spatial position ${\vec x}$ transverse to the motion of the electron at the time of closest approach),
see eq. (\ref{pug1}) or eq. (\ref{pglw}). Since
we have made the Lorentz transformation on the expression
$A^\mu(x)= e \frac{\beta^\mu_{\sim c}}{\beta_{\sim c} \cdot (x-X(\tau_0))}$
(see below), the Lorentz transformation technique we have used is valid at all the
time-space points $x^\mu$ to obtain the expression of the Coulomb potential
$A_0(x)$ produced by the electron at rest from the expression of the electromagnetic
potential $A^{\mu}(x)$ produced by the electron in uniform motion at its highest speed
(which is arbitrarily close to the speed of light $v\sim c$).

From eqs. (\ref{usc1a}), (\ref{ghf}) and (\ref{pglw}) we find that
\bea
{A^\mu}({x}) =e\frac{u^\mu_{\sim c}}{u_{\sim c} \cdot (x-X(\tau_0))}.
\label{eyspg1}
\eea
The Lorentz transformation of the electromagnetic potential $A^\mu(x)$ is given by \cite{jackson}
\bea
A'^{\mu } = \frac{\partial x'^\mu}{\partial x^\nu} A^\nu(x),~~~~~~~~~~~~~~~~~~~~~A'_\mu = \frac{\partial x^\nu}{\partial x'^\mu} A_\nu(x).
\label{ey1}
\eea

Consider two inertial reference frames $K$ and $K'$ with a relative velocity ${\vec v}$ between them. The time
and space coordinates of a point are $(t,~x,~y,~z)$ and $(t',~x',~y',~z')$ in the frames $K$ and $K'$
respectively. If the axes in two inertial frames $K$ and $K'$ remain parallel, but the uniform velocity
${\vec v}$ of the frame $K'$ in frame $K$ is in an arbitrary direction we find \cite{jackson}
\bea
x'_0 = \gamma \beta \cdot x,~~~~~~~~~~~~~~~~~~~~x'^i=x^i+\frac{\gamma -1}{{\vec {\beta}}^2} ({\vec { \beta}} \cdot {\vec x}) {\beta}^i-\gamma \beta^i x_0,~~~~~~~~~~i=1,2,3.
\label{e13cx}
\eea

From eqs. (\ref{e13cx}) and (\ref{bsc1}) we find
\bea
x'_0 = \gamma_{\sim c} (\beta_{\sim c} \cdot x),~~~~~~~~~~~~~~~~~~~~x'^i=x^i+\frac{\gamma_{\sim c} -1}{{\vec {\beta}}^2_{\sim c}} ({\vec { \beta}}_{\sim c} \cdot {\vec x}) {\beta}^i_{\sim c}-\gamma_{\sim c} \beta^i_{\sim c} x_0,~~~~~~~~~~i=1,2,3.\nonumber \\
\label{eyc13}
\eea
Hence from eqs. (\ref{ghf}) and (\ref{eyc13}) we find
\bea
x'_0 \neq \infty,~~~~~~~~~~~~~~~~~~~~x'^i \neq \infty,~~~~~~~~~~~~~~~~~{\rm when }~~~~~~~~~~~~~~~~v \sim c.
\label{eiyc13}
\eea
Since $x'_0$ and $x'^i$ are not exactly infinity (see eq. (\ref{eiyc13})) and since
the electromagnetic potential $A^{\mu }(x)$ in eq. (\ref{eyspg1}) or in eq. (\ref{a5ixc})
is neither zero nor (exactly) infinity [because electron has finite mass even if it is very small]
we can perform Lorentz transformation on $A^{\mu }(x)$ in eq. (\ref{eyspg1}) or on $A^{\mu }(x)$ in
eq. (\ref{a5ixc}).

From eq. (\ref{ey1}) we find
\bea
A'_0 = \frac{\partial x'_0}{\partial x^\nu} A^{\nu }(x).
\label{ey10}
\eea
From eqs. (\ref{ey10}), ({\ref{eyc13}) and (\ref{eyspg1}) [or (\ref{a5ixc})] we find
\bea
A'_0 = \gamma_{\sim c} \frac{\partial (\beta_{\sim c} \cdot x) }{\partial x^\nu} A^{\nu }(x) = \gamma_{\sim c} \times \beta_{\sim c} \cdot A(x)=ec\frac{1}{u_{\sim c} \cdot (x-X(\tau_0))}.
\label{eyc16}
\eea
Since electron has non-zero mass we find from eq. (\ref{ghf}) that the
$A'_0 \neq 0$ or  $A'_0 \neq \infty$ in eq. (\ref{eyc16}). Since
\bea
\tau = \tau'
\label{eyttpr}
\eea
under a Lorentz transformation, we find from eqs. (\ref{eyc13}) and (\ref{eyc16}) that
\bea
A'_0 =e \frac{1}{x'_0-X'_0(\tau'_0)}.
\label{eyc17}
\eea

By using the retarded condition
\bea
x'_0-X'_0(\tau'_0)=|{\vec x}' -{\vec X}'(\tau_0')|
\label{eyc17xx}
\eea
we find from eq. (\ref{eyc17})
\bea
A'_0 = e\frac{1}{|{\vec x}' -{\vec X}'(\tau_0')|}.
\label{eyc17xy}
\eea

Hence we find from eqs. (\ref{eyttpr}) and (\ref{eyc17xy})
\bea
A_0 = e\frac{1}{|{\vec x} -{\vec X}(\tau_0)|}
\label{eyc18xy}
\eea
which reproduces the exact expression of the Coulomb potential produced by the electron
at rest.

Hence we find that the exact expression of the Coulomb potential in eq. (\ref{eyc18xy})
can be obtained from the expression of the U(1) (approximate) pure gauge potential
in eq. (\ref{pglw}) [or from the expression of the electromagnetic potential $A^\mu(x)$
in eq. (\ref{a5ixc}) produced by the electron at its highest speed (which is arbitrarily
close to the speed of light $v\sim c$)] by using Lorentz transformations. Hence by
using Lorentz transformations we have proved that the static  Coulomb potential
$A_0(x)$ in eq. (\ref{colf}) can be obtained from the expression of the U(1) (approximate)
pure gauge potential $A^\mu(x)$ in eq. (\ref{pglw}) [or from the expression of the electromagnetic
potential $A^\mu(x)$ in eq. (\ref{a5ixc}) produced by the electron at its highest speed (which is
arbitrarily close to the speed of light $v\sim c$)] by using eq. (\ref{my2})
in (\ref{pglw}) [or by using eq. (\ref{my2}) in (\ref{a5ixc})].

From eq. (\ref{eyc18xy}) we find that the exact expression of the electromagnetic
potential $A^\mu(x)$ produced by the electron moving with arbitrary four velocity
$u^\mu(\tau)$ is given by
\bea
A^\mu({x}) =e\frac{u^\mu(\tau_0)}{u(\tau_0) \cdot (x-X(\tau_0))}
\label{etpg1}
\eea
which reproduces eq. (\ref{lw}) for the Lienard-Wiechert potential produced by the electron
moving with arbitrary four-velocity $u^\mu(\tau)$ where $\tau_0$ is obtained from the solution of the
retarded condition given by eq. (\ref{mx7}).

Similarly one can perform the Lorentz transformation on the electromagnetic field
$F^{\mu \nu}(x)$ produced by the electron in uniform motion at its highest speed
(which is arbitrarily close to the speed of light $v \sim c$) to obtain the exact
expression of the static Coulomb electric field produced by the electron at rest (see appendix B).

Hence one finds that once the expression of the (approximate) pure gauge electromagnetic potential
is known then one can obtain the exact expression of the static Coulomb potential from it by using
Lorentz transformations. This is consistent with the fact that eq. (\ref{colf}) can be
obtained from eq. (\ref{pglw}) by using eq. (\ref{my2}) in (\ref{pglw}).

Note that the above Lorentz transformation is performed in a frame which is moving arbitrarily
close to the speed of light ($v \sim c$) but not exactly at the speed of light ($v=c$). Since
the Lorentz transformation on the light-cone remains on the light-cone, one should not make
Lorentz transformation at speed of light ($v=c$). However, as shown above, one can make Lorentz
transformation in a frame which is moving arbitrarily close to the speed of light ($v \sim c$).
This is consistent with the fact that the electron has non-zero mass.

The above Lorentz transformation techniques can be extended to Yang-Mills theory.
From eqs. (\ref{usc1a}), (\ref{ghf}) and (\ref{spg1}) we find that a quark
in uniform motion at its highest speed (which is arbitrarily close to the speed of
light $v \sim c$) produces the Yang-Mills potential (color potential)
\bea
{A^\mu}^a({x}) =q^b(\tau_0)\frac{u^\mu_{\sim c}}{u_{\sim c} \cdot (x-X(\tau_0))}\left[\frac{{\rm exp}[g \int dl_c \frac{Q(\tau_0)}{l_c}]
-1}{g \int dl_c \frac{Q(\tau_0)}{l_c}}\right]_{ab}
\label{yspg1}
\eea
which can be written in the SU(3) (approximate) pure gauge potential form
(see eqs. (\ref{pure}), (\ref{su3pure}), (\ref{spg1}))
\bea
T^aA^{\mu a}(x) \sim \frac{1}{ig} [\partial^\mu U(x)]U^{-1}(x),~~~~~~~~~~~~~~U(x)=e^{igT^a\omega^a(x)}
\label{ylpgp}
\eea
where $dl_c$ integration is an indefinite integration and (see eq. (\ref{spg1}))
\bea
l_c=u_{\sim c} \cdot (x-X(\tau_0)),~~~~~~~Q_{ab}(\tau_0)=f^{abd}q^d(\tau_0),~~~~~~~\omega^a(x) = \int dl_c~ \frac{q^a(\tau_0)}{l_c}.
\label{ysqg1}
\eea

It has to be remembered that, as discussed explicitly in the section IV,
the expression $A^{\mu a}(x) =  \frac{\beta^\mu_{\sim c}}{\beta_{\sim c} \cdot
(x-X(\tau_0))}q^b(\tau_0) [\frac{e^{g\int dl_c \frac{Q(\tau_0)}{l_c}}-1}{g\int dl_c
\frac{Q(\tau_0)}{l_c}}]_{ab}$ in
eq. (\ref{afnab}) which is valid at all the time-space points $x^\mu$ can be written in the form of the SU(3)
(approximate) pure gauge potential at all the time-space points
$x^\mu$ (except at the spatial position ${\vec x}$ transverse to the motion of the quark at the time of closest approach),
see eq. (\ref{spg1}) or eq. (\ref{yspg1}). Since we have made the Lorentz
transformation on the expression
$A^{\mu a}(x) =  \frac{\beta^\mu_{\sim c}}{\beta_{\sim c} \cdot (x-X(\tau_0))}q^b(\tau_0)
[\frac{e^{g\int dl_c \frac{Q(\tau_0)}{l_c}}-1}{g\int dl_c \frac{Q(\tau_0)}{l_c}}]_{ab}$
(see below), the Lorentz transformation technique we have used is valid at all the
time-space points $x^\mu$ to obtain the form of the Yang-Mills potential
$A_0^a(x)$ produced by the quark at rest from the expression of the Yang-Mills
potential $A^{\mu a}(x)$ produced by the quark in uniform motion at its highest speed
(which is arbitrarily close to the speed of light $v\sim c$).

Since a quark has non-zero mass one finds from eq. (\ref{eiyc13}) that $x'_0$ and $x'^i$ are not exactly infinity.
Since $x'_0$ and $x'^i$ are not exactly infinity and the Yang-Mills potential $A^{\mu a}(x)$ in eq. (\ref{yspg1})
or in eq. (\ref{afnab}) is neither zero nor (exactly) infinity [because a quark has finite mass even if the light
quark mass is very small] we can perform Lorentz transformation on $A^{\mu a}(x)$ in eq. (\ref{yspg1}) or on
$A^{\mu a}(x)$ in eq. (\ref{afnab}).

The Lorentz transformation of the Yang-Mills potential $A^{\mu a}(x)$ is given by \cite{jackson}
\bea
A'^{\mu a} = \frac{\partial x'^\mu}{\partial x^\nu} A^{\nu a}(x),~~~~~~~~~~~~~~~~~~~~~A'^a_\mu(x') = \frac{\partial x^\nu}{\partial x'^\mu} A^a_\nu(x).
\label{y1}
\eea
From eq. (\ref{y1}) we find
\bea
A'^a_0 = \frac{\partial x'_0}{\partial x^\nu} A^{\nu a}(x).
\label{y10}
\eea
From eqs. (\ref{y10}), ({\ref{eyc13}) and (\ref{yspg1}) [or (\ref{afnab})] we find
\bea
A'^a_0 = \gamma_{\sim c} \frac{\partial (\beta_{\sim c} \cdot x) }{\partial x^\nu} A^{\nu a}(x) = \gamma_{\sim c} \times \beta_{\sim c} \cdot A^a(x)=c\frac{q^b(\tau_0)}{u_{\sim c} \cdot (x-X(\tau_0))}\left[\frac{{\rm exp}[g \int dl_c \frac{Q(\tau_0)}{l_c}]
-1}{g \int dl_c \frac{Q(\tau_0)}{l_c}}\right]_{ab}.\nonumber \\
\label{yc16}
\eea
Since a quark has non-zero mass we find from eq. (\ref{ghf}) that the
$A'^a_0 \neq 0$ or $A'^a_0 \neq \infty$ in eq. (\ref{yc16}). Hence we find from eqs. (\ref{eyc13}),
(\ref{eyttpr}) and (\ref{yc16}) that
\bea
A'^a_0 = \frac{q^b(\tau'_0)}{x'_0-X'_0(\tau'_0)}\left[\frac{{\rm exp}[g \int dl'_0 \frac{Q(\tau'_0)}{l'_0}]
-1}{g \int dl'_0 \frac{Q(\tau'_0)}{l'_0}}\right]_{ab}
\label{yc17}
\eea
where
\bea
l'_0=x'_0-X'_0(\tau'_0).
\label{ysqg2}
\eea

By using the retarded condition
\bea
x'_0-X'_0(\tau'_0)=|{\vec x}' -{\vec X}'(\tau_0')|
\label{yc17xx}
\eea
we find from eq. (\ref{yc17})
\bea
A'^a_0 = \frac{q^b(\tau'_0)}{|{\vec x}' -{\vec X}'(\tau_0')|}\left[\frac{{\rm exp}[g \int dl' \frac{Q(\tau'_0)}{l'}]
-1}{g \int dl' \frac{Q(\tau'_0)}{l'}}\right]_{ab}
\label{yc17xy}
\eea
where
\bea
l'=|{\vec x}'-{\vec X}'(\tau'_0)|.
\label{ysqg3}
\eea

Hence we find from eqs. (\ref{eyttpr}) and (\ref{yc17xy})
\bea
A^a_0 = \frac{q^b(\tau_0)}{|{\vec x}-{\vec X}(\tau_0)|}\left[\frac{{\rm exp}[g \int dr \frac{Q(\tau_0)}{r}]
-1}{g \int dr \frac{Q(\tau_0)}{r}}\right]_{ab}
\label{yc18xy}
\eea
which is the general expresion of the color potential (Yang-Mills potential) produced by the color charges $q^a(\tau)$
of the quark at rest where $dr$ is an indefinite integration, ${\vec X}(\tau_0)$ is the position vector of
the quark at the retarded time and
\bea
r=|{\vec x}-{\vec X}(\tau_0)|,~~~~~~~~~~~~~~~~~~~Q_{ab}(\tau_0)=f^{abd}q^d(\tau_0).
\label{ysqg4}
\eea
In eqs. (\ref{yc18xy}) and (\ref{ysqg4}) the repeated indices $b,d=(1,2,...8)$ are summed.

Eq. (\ref{yc18xy}) in Yang-Mills theory is analogous to eq. (\ref{eyc18xy}) in Maxwell theory.

From eq. (\ref{yc18xy}) we find that the general expression of the color potential (Yang-Mills
potential) $A^{\mu a}(x)$ produced by the color charges $q^a(\tau)$ of the quark
moving with arbitrary four velocity $u^\mu(\tau)$ is given by
\bea
A^{\mu a}({x}) =q^b(\tau_0)\frac{u^\mu(\tau_0)}{u(\tau_0) \cdot (x-X(\tau_0))}\left[\frac{{\rm exp}[g \int dl \frac{Q(\tau_0)}{l}]
-1}{g \int dl \frac{Q(\tau_0)}{l}}\right]_{ab}
\label{tpg1}
\eea
where $dl$ integration is an indefinite integration,
\bea
l=u(\tau_0) \cdot (x-X(\tau_0)),~~~~~~~~~~~~~~~~~~~~~~~~~Q_{ab}(\tau_0)=f^{abc}q^c(\tau_0),
\label{tqg1}
\eea
the repeated color indices $b,c$(=1,2,...8) are summed and $\tau_0$ is determined from the solution of the
retarded condition given by eq. (\ref{mx7}).

Eq. (\ref{tpg1}) in Yang-Mills theory is analogous to eq. (\ref{etpg1}) in Maxwell theory.

Note that unlike Maxwell theory, the Yang-Mills color current density generated by the Yang-Mills
potential in eq. (\ref{tpg1}) is non-linear function of color charges $q^a(\tau)$ of the quark,
see appendix C.

From the expression of the proper time
\bea
\tau = \int \frac{dX_0(\tau)}{c\gamma(X_0(\tau))}
\eea
we find that when ${\vec \beta}=0$
\bea
\tau_0=\frac{X_0(\tau_0)}{c}.
\label{q1}
\eea
From the retarded condition from eq. (\ref{mx7}) we find
\bea
x_0-X_0(\tau_0)=ct-X_0(\tau_0)=r
\label{q2}
\eea
where
\bea
r=|{\vec x} -{\vec X}(\tau_0)|.
\label{ojmb}
\eea
From eqs. (\ref{q1}) and (\ref{q2}) we find
\bea
\tau_0 = t-\frac{r}{c}.
\label{q3}
\eea

Using eq. (\ref{q3}) in (\ref{yc18xy}) we find
\bea
\Phi^a(x)=A_0^a(t,{\bf x}) =\frac{q^b(t-\frac{r}{c})}{r}\left[\frac{{\rm exp}[g\int dr \frac{Q(t-\frac{r}{c})}{r}]
-1}{g \int dr \frac{Q(t-\frac{r}{c})}{r}}\right]_{ab}
\label{upg1fg}
\eea
which reproduces eq. (\ref{upb}) where $dr$ integration is an indefinite integration, ${\vec X}(\tau_0)$ is the position
vector of the quark at the retarded time,
\bea
r=|{\vec x}-{\vec X}(\tau_0)|,~~~~~~~~~~~~~~~~Q_{ab}(\tau_0)=f^{abd}q^d(\tau_0)
\label{uqg1}
\eea
and the repeated color indices $b,d$(=1,2,...8) are summed.

The eq. (\ref{upg1fg}) in Yang-Mills theory is analogous to Coulomb potential $\Phi(x)=A_0({\bf x})=\frac{e}{r}$
in Maxwell theory.

We find from eq. (\ref{upg1fg}) that, unlike Coulomb potential $A_0({\bf x})=\frac{e}{r}$
produced by the electric charge of the electron at rest in Maxwell theory which is independent of time $t$, the color
potential $A_0^a(t,{\bf x})$ produced by the color charges of the quark in Yang-Mills theory depends on time $t$ even
if the quark is at rest. This is a consequence of time dependent color charges of the quark in Yang-Mills theory. The
color potential (Yang-Mills potential) $A_0^a(x)$ at time $t$ depends on color charges of the quark at the retarded time
$t-\frac{r}{c}$.

Note that when the color charge $q^a$ is constant we find from eq. (\ref{upg1fg})
\bea
\Phi^a(x) =\frac{q^a}{r}
\label{ffg}
\eea
which reproduces the Coulomb-like potential, similar to Maxwell theory
where the constant electric charge $e$ produces Coulomb potential
$\Phi(x)=\frac{e}{r}$.

\section{General Form of the Color Charge of the Quark}

The exact form of the color potential produced by color charges of the quark may
provide an insight to the question "why quarks are confined inside a (stable) proton".
Hence in order to find the exact form of the color potential produced by the color charges of the quark we
need to find the exact form of eight time dependent fundamental color charges $q^a(t)$ of the quark where
$a=1,2,...8$ are color indices. The general form of the eight time dependent fundamental
color charges $q^a(t)$ of the quark in Yang-Mills theory in SU(3) is defined in \cite{nayak2} which
we will briefly describe here.

As we have explicitly described in the introduction, the Yang-Mills theory was developed by making analogy with the
Maxwell theory by extending the U(1) group to SU(3) group appropriately (see \cite{yang}). Hence the
analogy with the Maxwell theory plays a crucial role to predict many quantities in Yang-Mills
theory. For example we saw in sections III and V that the expression of the electromagnetic
potential (Lienard-Weichert potential) $A^\mu(x)$ produced by the electric charge $e$ of the
electron in Maxwell theory can be obtained from the expression of the U(1) pure gauge potential produced by
the electron. Hence in analogy with Maxwell theory we have found in sections IV and V that the general expression of the
Yang-Mills potential (color potential) $A^{\mu a}(x)$ produced by the color charges $q^a(\tau)$ of the quark
in Yang-Mills theory can be obtained from the general expression of the SU(3) pure gauge potential produced by
the quark.

Similarly the analogy with the Maxwell theory plays a crucial role to predict the general form of the
fundamental time dependent color charge $q^a(t)$ of the quark from the Dirac wave function $\psi_i(x)$
of the quark.

First of all we observe that in Maxwell theory (U(1) gauge theory) the form of the fundamental electric
charge $e$ of the electron can be obtained from the Dirac wave function $\psi(x)$ of the electron by
using, 1) the continuity equation $\partial_\mu j^\mu(x) = 0$ of the electric current density $j^\mu(x)$ of the
electron, 2) scalar component $j_0(x) = e{\bar \psi}(x) \gamma_0 \psi(x)= e\psi^\dagger(x) \psi(x)$
of the electric current density $j^\mu(x) = e{\bar \psi}(x) \gamma^\mu \psi(x)$ of the electron and
3) number density $n_e(x)=\psi^\dagger(x) \psi(x)$ of the electron by using the normalization condition
$\int d^3x \psi^\dagger(x) \psi(x)=1$. Hence in analogy to Maxwell theory (U(1) gauge theory), we find
that the form of the color charge $q^a(t)$ of the quark in Yang-Mills theory (SU(3) gauge theory)
can be obtained from the Dirac wave function $\psi_i(x)$ of the quark
by using, 1) the equation $D_\mu[A] j^{\mu a}(x) = 0$ of the color current density $j^{\mu a}(x)$ of the quark,
2) scalar component $j_0^a(x) = g{\bar \psi}_i(x) \gamma_0 T^a_{ik}\psi_k(x)= g\psi^\dagger_i(x) T^a_{ik}\psi_k(x)$
of the color current density $j^{\mu a}(x) = g{\bar \psi}_i(x) \gamma^\mu T^a_{ik} \psi_k(x)$ of the quark
and 3) number density $n_q(x)=\psi^\dagger_i(x) \psi_i(x)$ of the quark by using the normalization condition
$\int d^3x \psi^\dagger_i(x) \psi_i(x)=1$ \cite{nayak2}.

Hence by using the analogy with the Maxwell theory (U(1) gauge theory) we find that the general form of eight
fundamental time dependent color charges $q^a(t)$ of the quark in Yang-Mills theory in SU(3) is given by
\cite{nayak2}
\bea
&& q_1(t) = g\times {\rm sin}\theta(t) \times {\rm sin}\sigma(t)\times {\rm cos}\eta(t)\times {\rm cos}\phi_{12}(t), \nonumber \\
&& q_2(t) = g\times {\rm sin}\theta(t) \times {\rm sin}\sigma(t)\times {\rm cos}\eta(t)\times {\rm sin}\phi_{12}(t), \nonumber \\
&& q_3(t) = g\times {\rm cos}\theta(t) \times {\rm sin}\phi(t) \nonumber \\
&& q_4(t) = g\times {\rm sin}\theta(t) \times {\rm sin}\sigma(t)\times {\rm sin}\eta(t)\times {\rm cos}\phi_{13}(t), \nonumber \\
&& q_5(t) = g\times {\rm sin}\theta(t) \times {\rm sin}\sigma(t)\times {\rm sin}\eta(t)\times {\rm sin}\phi_{13}(t), \nonumber \\
&& q_6(t) = g\times {\rm sin}\theta(t)\times {\rm cos}\sigma(t) \times {\rm cos}\phi_{23}(t), \nonumber \\
&& q_7(t) = g\times {\rm sin}\theta(t)\times {\rm cos}\sigma(t)\times {\rm sin}\phi_{23}(t), \nonumber \\
&& q_8(t) = g\times {\rm cos}\theta(t)\times {\rm cos}\phi(t)
\label{qspin}
\eea
where the ranges of the time dependent phases are given by
\bea
&&{\rm sin}^{-1}(\sqrt{\frac{2}{3}}) ~~\le ~~ \theta(t) ~~ \le ~~\pi-{\rm sin}^{-1}(\sqrt{\frac{2}{3}}),~~~~~~~~~~~0 \le \sigma(t),~\eta(t) \le \frac{\pi}{2},\nonumber \\
&& 0~\le ~ \phi(t)\le 2 \pi,~~~~~~~~~~~~~~~~~~~~~~~~~~~~~~~~-\pi < \phi_{12}(t),~\phi_{13}(t),~\phi_{23}(t) \le \pi.
\label{3ina}
\eea
Note that all the real time dependent phases
$\theta(t),~\sigma(t),~\eta(t),~\phi(t),~\phi_{12}(t),~\phi_{13}(t),~\phi_{23}(t)$
in eq. (\ref{qspin}) can not be independent of time $t$ because if all of them are
independent of $t$ then the Yang-Mills potential $A_0^a(x)$ in eq. (\ref{upb}) reduces to
Coulomb-like potential $A_0^a(x)=\frac{q^a}{r}$. Since Coulomb-like potential
$A_0^a(x)=\frac{q^a}{r}$ can not explain confinement of quarks inside (stable)
proton one finds that the real phases
$\theta(t),~\sigma(t),~\eta(t),~\phi(t),~\phi_{12}(t),~\phi_{13}(t),~\phi_{23}(t)$
in eq. (\ref{qspin}) have to be time dependent.

It can be seen that the general form of the eight fundamental time dependent color charges $q^a(t)$
of the quark in eq. (\ref{qspin}) depend on the universal coupling $g$ which is a fundamental quantity
that appears in the Yang-Mills Lagrangian density \cite{yang,mutta}.

Since the fundamental time dependent color charge $q^a(t)$ of the quark
in eq. (\ref{qspin}) is linearly proportional to $g$ and the Yang-Mills potential $A^{\mu a}(x)$ in
eq. (\ref{tpg1}) contains infinite powers of $g$ we find that the definition of the fundamental
time dependent color charge $q^a(t)$ of the quark in eq. (\ref{qspin}) is independent of the Yang-Mills
potential $A^{\mu a}(x)$.

\section{Conclusion}

Constant electric charge $e$ satisfies the continuity equation $\partial_\mu j^{\mu }(x)= 0$ where
$j^\mu(x)$ is the current density of the electron. However, the Yang-Mills color current density
$j^{\mu a}(x)$ of the quark satisfies the equation $D_\mu[A] j^{\mu a}(x)= 0$ which is not a continuity equation
($\partial_\mu j^{\mu a}(x)\neq 0$) which implies that a color charge $q^a(t)$ of the quark is not constant but it is
time dependent where $a=1,2,...8$ are color indices. In this paper
we have derived general form of color potential produced by color charges of the quark.
We have found that the general form of the color potential produced by the color charges
of the quark at rest is given by
$\Phi^a(x) =A_0^a(t,{\bf x}) =\frac{q^b(t-\frac{r}{c})}{r}\left[\frac{{\rm exp}[g\int dr \frac{Q(t-\frac{r}{c})}{r}]
-1}{g \int dr \frac{Q(t-\frac{r}{c})}{r}}\right]_{ab}$ where $dr$ integration is an indefinite integration, ~~
$Q_{ab}(\tau_0)=f^{abd}q^d(\tau_0)$, ~~$r=|{\vec x}-{\vec X}(\tau_0)|$, ~~$\tau_0=t-\frac{r}{c}$
is the retarded time, ~~$c$ is the speed of light,
~~${\vec X}(\tau_0)$ is the position of the quark at the retarded time and the repeated color indices $b,d$(=1,2,...8) are summed.
For constant color charge $q^a$ we reproduce the Coulomb-like potential $\Phi^a(x)=\frac{q^a}{r}$
which is consistent with the Maxwell theory where constant electric charge $e$ produces the Coulomb potential
$\Phi(x)=\frac{e}{r}$.

\acknowledgments

I thank George Sterman for useful discussions, in particular, with respect to Lorentz transformation
near speed of light. I also thank Robert Shrock for discussions on lattice QCD.

\appendix

\section{ Derivation of the General Expression of $\omega^a(x)$ of the Quark}

In this appendix we will present the mathematical details which are used in section IV
to derive the general expression of $\omega^a(x)$ of the quark. From eqs. (\ref{mx6}),
(\ref{4va}), (\ref{dtau0}) and (\ref{dvu}) we find that
\bea
&&\partial_\nu {\cal A}^{\mu a}(x) = q^a\frac{ (x-X(\tau_0))_\nu {\dot u}^\mu(\tau_0)}{[u(\tau_0) \cdot (x-X(\tau_0))]^2}\nonumber \\
&&-q^a\frac{ u^\mu(\tau_0)}{[u(\tau_0) \cdot (x-X(\tau_0))]^2}[\frac{[\dot{u}(\tau_0) \cdot (x-X(\tau_0))-c^2](x-X(\tau_0))_\nu }{u(\tau_0) \cdot (x-X(\tau_0))}+u_\nu(\tau_0)]
\label{mx8c}
\eea
where ${\dot u}^\mu(\tau_0)$ is given by eq. (\ref{ttp}). Eq. (\ref{mx8c}) gives
\bea
&&~\partial_\mu {\cal A}^{\mu a}(x) = q^a\frac{ (x-X(\tau_0))_\mu {\dot u}^\mu(\tau_0)}{[u(\tau_0) \cdot (x-X(\tau_0))]^2}\nonumber \\
&&-q^a\frac{ u^\mu(\tau_0)}{[u(\tau_0) \cdot (x-X(\tau_0))]^2}[\frac{[\dot{u}(\tau_0) \cdot (x-X(\tau_0))-c^2](x-X(\tau_0))_\mu }{u(\tau_0) \cdot (x-X(\tau_0))}+u_\mu(\tau_0)]
\label{mx8a}
\eea
which is used to derive eq. (\ref{mx8b}).

From eqs. (\ref{mx8c}), (\ref{4va}), (\ref{dtau0}) and (\ref{dvu}) we find
\bea
&&~\partial^\nu \partial_\nu {\cal A}^{\mu a}(x) = q^a\frac{\partial^\nu [(x-X(\tau_0))_\nu {\dot u}^\mu(\tau_0)]}{[u(\tau_0) \cdot (x-X(\tau_0))]^2}-2q^a\frac{(x-X(\tau_0))_\nu {\dot u}^\mu(\tau_0)}{[u(\tau_0) \cdot (x-X(\tau_0))]^3}u^\nu(\tau_0)\nonumber \\
&&-q^a\frac{ (x-X(\tau_0))^\nu {\dot u}^\mu(\tau_0)}{[u(\tau_0) \cdot (x-X(\tau_0))]^3}u_\nu(\tau_0)+2q^a\frac{ u^\mu(\tau_0)}{[u(\tau_0) \cdot (x-X(\tau_0))]^3}[2\dot{u}(\tau_0) \cdot (x-X(\tau_0))-c^2]\nonumber \\
&&-q^a\frac{ u^\mu(\tau_0)}{[u(\tau_0) \cdot (x-X(\tau_0))]^2}[\frac{[\dot{u}(\tau_0) \cdot \partial^\nu (x-X(\tau_0))](x-X(\tau_0))_\nu }{u(\tau_0) \cdot (x-X(\tau_0))}]\nonumber \\
&&-3q^a\frac{ u^\mu(\tau_0)}{[u(\tau_0) \cdot (x-X(\tau_0))]^2}[\frac{[\dot{u}(\tau_0) \cdot (x-X(\tau_0))-c^2] }{u(\tau_0) \cdot (x-X(\tau_0))}]\nonumber \\
&&+q^a\frac{ u^\mu(\tau_0)}{[u(\tau_0) \cdot (x-X(\tau_0))]^3}[{\dot u}(\tau_0) \cdot (x-X(\tau_0))-c^2]-q^a\frac{ u^\mu(\tau_0)}{[u(\tau_0) \cdot (x-X(\tau_0))]^3}[{\dot u}(\tau_0) \cdot (x-X(\tau_0))]\nonumber \\
\eea
which gives
\bea
&&~\partial^\nu \partial_\nu {\cal A}^{\mu a}(x) = 3q^a\frac{{\dot u}^\mu(\tau_0)}{[u(\tau_0) \cdot (x-X(\tau_0))]^2}-2q^a\frac{ {\dot u}^\mu(\tau_0)}{[u(\tau_0) \cdot (x-X(\tau_0))]^2}-q^a\frac{ {\dot u}^\mu(\tau_0)}{[u(\tau_0) \cdot (x-X(\tau_0))]^2}\nonumber \\
&&+q^a\frac{ u^\mu(\tau_0)}{[u(\tau_0) \cdot (x-X(\tau_0))]^3}[\dot{u}(\tau_0) \cdot (x-X(\tau_0))+c^2]-q^a\frac{ u^\mu(\tau_0)}{[u(\tau_0) \cdot (x-X(\tau_0))]^3}[\dot{u}(\tau_0) \cdot (x-X(\tau_0))]\nonumber \\
&&+q^a\frac{ u^\mu(\tau_0)}{[u(\tau_0) \cdot (x-X(\tau_0))]^3}[{\dot u}(\tau_0) \cdot (x-X(\tau_0))-c^2]-q^a\frac{ u^\mu(\tau_0)}{[u(\tau_0) \cdot (x-X(\tau_0))]^3}[{\dot u}(\tau_0) \cdot (x-X(\tau_0))]\nonumber \\
\label{mx8d}
\eea
which is used to derive eq. (\ref{mx8e}).

From eqs. (\ref{mx10}), (\ref{4va}), (\ref{dtau0}) and (\ref{dvu}) we find that
\bea
&&\partial_\nu {\cal A}^{\mu a}(x) = q^a(\tau_0)\frac{ (x-X(\tau_0))_\nu {\dot u}^\mu(\tau_0)}{[u(\tau_0) \cdot (x-X(\tau_0))]^2}-q^a(\tau_0)\frac{ u^\mu(\tau_0)}{[u(\tau_0) \cdot (x-X(\tau_0))]^2}\nonumber \\
&&~[\frac{[\dot{u}(\tau_0) \cdot (x-X(\tau_0))-c^2](x-X(\tau_0))_\nu }{u(\tau_0) \cdot (x-X(\tau_0))}+u_\nu(\tau_0)]+[\partial^\nu q^a(\tau_0)] \frac{ u_\mu(\tau_0)}{u(\tau_0) \cdot (x-X(\tau_0))}. \nonumber \\
\label{mxb}
\eea

From eqs. (\ref{mxb}), (\ref{4va}), (\ref{dtau0}) and (\ref{dvu}) we find
\bea
&&\partial^\nu \partial_\nu {\cal A}^{\mu a}(x) =[\partial^\nu q^a(\tau_0)]\frac{ (x-X(\tau_0))_\nu {\dot u}^\mu(\tau_0)}{[u(\tau_0) \cdot (x-X(\tau_0))]^2}-[\partial^\nu q^a(\tau_0)]\frac{ u^\mu(\tau_0)}{[u(\tau_0) \cdot (x-X(\tau_0))]^2}\nonumber \\
&&~[\frac{[\dot{u}(\tau_0) \cdot (x-X(\tau_0))-c^2](x-X(\tau_0))_\nu }{u(\tau_0) \cdot (x-X(\tau_0))}+u_\nu(\tau_0)]+[\partial_\nu q^a(\tau_0)] [\partial^\nu [\frac{ u^\mu(\tau_0)}{u(\tau_0) \cdot (x-X(\tau_0))}]]\nonumber \\
&&+[\partial^\nu \partial^\nu q^a(\tau_0)] \frac{ u^\mu(\tau_0)}{u(\tau_0) \cdot (x-X(\tau_0))}
\eea
which gives by using eqs. (\ref{4va}), (\ref{dtau0}) and (\ref{dvu})
\bea
&&\partial^\nu \partial_\nu {\cal A}^{\mu a}(x) =-{\dot q}^a(\tau_0)\frac{ u^\mu(\tau_0)}{[u(\tau_0) \cdot (x-X(\tau_0))]^2}-[\frac{ (x-X(\tau_0))^\nu {\dot q}^a(\tau_0)}{u(\tau_0) \cdot (x-X(\tau_0))}] \nonumber \\
&&~[\frac{ u^\mu(\tau_0)}{[u(\tau_0) \cdot (x-X(\tau_0))]^2}] [\frac{[\dot{u}(\tau_0) \cdot (x-X(\tau_0))-c^2](x-X(\tau_0))_\nu }{u(\tau_0) \cdot (x-X(\tau_0))}+u_\nu(\tau_0)] \nonumber \\
&&+[\frac{ (x-X(\tau_0))_\nu (x-X(\tau_0))^\nu {\ddot q}^a(\tau_0)}{[u(\tau_0) \cdot (x-X(\tau_0))]^2}] \frac{ u^\mu(\tau_0)}{u(\tau_0) \cdot (x-X(\tau_0))}\nonumber \\
&&+[\frac{[\partial^\nu  (x-X(\tau_0))_\nu] {\dot q}^a(\tau_0)}{u(\tau_0) \cdot (x-X(\tau_0))}] \frac{ u^\mu(\tau_0)}{u(\tau_0) \cdot (x-X(\tau_0))}\nonumber \\
&&-[\frac{ (x-X(\tau_0))_\nu {\dot q}^a(\tau_0)}{[u(\tau_0) \cdot (x-X(\tau_0))]^2}] \frac{ u^\mu(\tau_0)}{u(\tau_0) \cdot (x-X(\tau_0))}[\frac{[\dot{u}(\tau_0) \cdot (x-X(\tau_0))-c^2](x-X(\tau_0))^\nu }{u(\tau_0) \cdot (x-X(\tau_0))}+u^\nu(\tau_0)] \nonumber \\
\label{mxc}
\eea
where ${\dot q}^a(\tau_0)$ is give by eq. (\ref{qqpi}). Simplifying eq. (\ref{mxc}) we find
\bea
&&\partial^\nu \partial_\nu {\cal A}^{\mu a}(x) =-{\dot q}^a(\tau_0)\frac{ u^\mu(\tau_0)}{[u(\tau_0) \cdot (x-X(\tau_0))]^2}- {\dot q}^a(\tau_0)
\frac{ u^\mu(\tau_0)}{[u(\tau_0) \cdot (x-X(\tau_0))]^2}\nonumber \\
&&+3{\dot q}^a(\tau_0) \frac{ u^\mu(\tau_0)}{[u(\tau_0) \cdot (x-X(\tau_0))]^2}-{\dot q}^a(\tau_0) \frac{ u^\mu(\tau_0)}{[u(\tau_0) \cdot (x-X(\tau_0))]^2}
\label{mxd}
\eea
which is used to derive eq. (\ref{mxe}).

From eqs. (\ref{mxb}) and (\ref{dtau0}) we find
\bea
&&\partial_\mu {\cal A}^{\mu a}(x) = q^a(\tau_0)\frac{ (x-X(\tau_0))_\mu {\dot u}^\mu(\tau_0)}{[u(\tau_0) \cdot (x-X(\tau_0))]^2}-q^a(\tau_0)\frac{ u^\mu(\tau_0)}{[u(\tau_0) \cdot (x-X(\tau_0))]^2}\nonumber \\
&&~[\frac{[\dot{u}(\tau_0) \cdot (x-X(\tau_0))-c^2](x-X(\tau_0))_\mu }{u(\tau_0) \cdot (x-X(\tau_0))}+u_\mu(\tau_0)]+{\dot q}^a(\tau_0) \frac{ (x-X(\tau_0))_\mu }{[u(\tau_0) \cdot (x-X(\tau_0))]^2}u^\mu(\tau_0)\nonumber \\
\label{mxa}
\eea
which is used to derive eq. (\ref{mxhi}).

From eqs. (\ref{mxb}) and (\ref{dtau0}) we find
\bea
&&\partial^\mu {\cal A}^{\nu a}(x)-\partial^\nu {\cal A}^{\mu a}(x) = q^a(\tau_0)\frac{(x-X(\tau_0))^\mu {\dot u}^\nu(\tau_0)- (x-X(\tau_0))^\nu {\dot u}^\mu(\tau_0)}{[u(\tau_0) \cdot (x-X(\tau_0))]^2}\nonumber \\
&&-q^a(\tau_0)~[\frac{[\dot{u}(\tau_0) \cdot (x-X(\tau_0))-c^2][(x-X(\tau_0))^\mu u^\nu(\tau_0)-(x-X(\tau_0))^\nu u^\mu(\tau_0)]}{[u(\tau_0) \cdot (x-X(\tau_0))]^3}]\nonumber \\
&&+{\dot q}^a(\tau_0) \frac{ (x-X(\tau_0))^\mu u^\nu(\tau_0)-(x-X(\tau_0))^\nu u^\mu(\tau_0)}{[u(\tau_0) \cdot (x-X(\tau_0))]^2}.
\label{mxn}
\eea
For uniform velocity we obtain from eq. (\ref{mxn})
\bea
&&\partial^\mu {\cal A}^{\nu a}(x)-\partial^\nu {\cal A}^{\mu a}(x) = q^a(\tau_0)~\frac{(x-X(\tau_0))^\mu u^\nu-(x-X(\tau_0))^\nu u^\mu}{[u \cdot (x-X(\tau_0))]^3}\nonumber \\
&&+{\dot q}^a(\tau_0) \frac{ (x-X(\tau_0))^\mu u^\nu-(x-X(\tau_0))^\nu u^\mu}{[u \cdot (x-X(\tau_0))]^2}.
\label{mxo}
\eea
When the quark in uniform motion at its highest speed (which is arbitrarily close to the speed of
light $v \sim c$, see eq. (\ref{bsc1}), we find from eq. (\ref{mxo})
\bea
&&\partial^\mu {\cal A}^{\nu a}(x)-\partial^\nu {\cal A}^{\mu a}(x)= q^a(\tau_0)~\frac{(x-X(\tau_0))^\mu u^\nu_{\sim c}-(x-X(\tau_0))^\nu u^\mu_{\sim c}}{[u_{\sim c} \cdot (x-X(\tau_0))]^3}\nonumber \\
&&+{\dot q}^a(\tau_0) \frac{ (x-X(\tau_0))^\mu u^\nu_{\sim c}-(x-X(\tau_0))^\nu u^\mu_{\sim c}}{[u_{\sim c} \cdot (x-X(\tau_0))]^2}
\label{mxp}
\eea
which is used to derive eq. (\ref{mxq}).

\section{ Lorentz Transformation of the Electromagnetic Field Produced by an Electron Moving Arbitrarily Close to the Speed of Light }

The electromagnetic field produced by the electron of charge $e$ in uniform motion
is given by eq. (\ref{fmnu}) where $\tau_0$ is defined from the the retarded condition
[see eq. (\ref{mx7})]
\bea
x_0 -X_0(\tau_0) = |{\vec x} - {\vec X}(\tau_0)|=R.
\label{7ixx}
\eea
The Lorentz transformation of the electromagnetic field is given by \cite{jackson}
\bea
F'^{\mu \nu} = \frac{\partial x'^\mu}{\partial x^\alpha}\frac{\partial x'^\nu}{\partial x^\beta} F^{\alpha \beta}(x),~~~~~~~~~~~~~~~F'_{\mu \nu}=\frac{\partial x^\alpha }{\partial x'^\mu} \frac{\partial x^\beta}{\partial x'^\nu} F_{\alpha \beta}(x).
\label{3}
\eea

For electron in uniform motion at its highest speed (which is arbitrarily close to the speed of
light $v \sim c$) we find from eqs. (\ref{fmnu}), (\ref{usc1a}) and (\ref{bsc1})
\bea
F^{\mu \nu}(x) = ec^2 \frac{(x-X(\tau_0))^\mu u^\nu_{\sim c}-(x-X(\tau_0))^\nu u^\mu_{\sim c}}{[u_{\sim c} \cdot (x-X(\tau_0))]^3}.
\label{c5ixx}
\eea

At the spatial position ${\vec x}$ transverse to the motion of the electron at the time of closest approach
we find from eq. (\ref{c5ixx}) that
\bea
F^{\mu \nu}(x)=ec^2 \frac{(x-X(\tau_0))^\mu u^\nu_{\sim c}-(x-X(\tau_0))^\nu u^\mu_{\sim c}}{[u_{\sim c} \cdot (x-X(\tau_0))]^3}\propto \gamma_{\sim c}
\label{pgfi}
\eea
and at all other time-space points $x^\mu$ we find from eq. (\ref{c5ixx}) that
\bea
F^{\mu \nu}(x)=ec^2 \frac{(x-X(\tau_0))^\mu u^\nu_{\sim c}-(x-X(\tau_0))^\nu u^\mu_{\sim c}}{[u_{\sim c} \cdot (x-X(\tau_0))]^3} \sim 0.
\label{pgf}
\eea

Hence from eqs. (\ref{c5ixx}), (\ref{pgfi}) and (\ref{pgf}) we find that when electron in uniform motion is
at its highest speed (which is arbitrarily close to the speed of light $v \sim c$), it produces
(approximate) pure gauge electromagnetic field at all the time-space points $x^\mu$ (except at the
spatial position ${\vec x}$ transverse to the motion of the electron at the time of closest approach).
We call it (approximate) pure gauge
electromagnetic field $F^{\mu \nu}(x)$ because electron has non-zero mass (even if very small) and hence it
can not travel exactly at the speed of light $v=c$ to produce exact pure gauge electromagnetic field
\bea
F^{\mu \nu }(x)=0.
\label{expg}
\eea
Hence for electron we find from eqs. (\ref{pgf}) and (\ref{c5ixx})
\bea
F^{\mu \nu}(x) \neq 0,~~~~~~~~~~~~{\rm when},~~~~~~~~~~~~~~~v\sim c,~~~~~~~~~~~~~~~~~~~{\rm even~if}~~~~~~~~~~~~~~~~F^{\mu \nu}(x) \sim 0.\nonumber \\
\label{nexpg}
\eea
Similarly from eqs. (\ref{pgfi}) and (\ref{c5ixx}) we find that 
\bea
F^{\mu \nu}(x) \neq \infty,~~~~~~~{\rm when},~~~~~~~~~v\sim c,~~~~~~~~~~~~~{\rm even~if}~~~~~~~~~~~~~~~~F^{\mu \nu}(x) \propto \gamma_{\sim c}.\nonumber \\
\label{inf}
\eea

Since $x'_0$ and $x'^i$ are not exactly infinity (see eq. (\ref{eiyc13})) and since
$F^{\mu \nu}(x)$ in eq. (\ref{pgf}) or in eq. (\ref{pgfi})
is neither (exactly) zero nor (exactly) infinity we can perform Lorentz transformation
on $F^{\mu \nu}(x)$ in eq. (\ref{c5ixx}). It has to be remembered that,
the expression $F^{\mu \nu}(x) = ec^2 \frac{(x-X(\tau_0))^\mu u^\nu_{\sim c}-(x-X(\tau_0))^\nu u^\mu_{\sim c}}{[u_{\sim c} \cdot (x-X(\tau_0))]^3}$
in eq. (\ref{c5ixx}) which is valid at all the time-space points $x^\mu$
can be written in the form of the U(1) (approximate) pure gauge electromagnetic field
at all the time-space points
$x^\mu$ (except at the spatial position ${\vec x}$ transverse to the motion of the electron at the time of closest approach),
see eq. (\ref{pgf}). Since we have made the Lorentz transformation on the expression
$F^{\mu \nu}(x) = ec^2 \frac{(x-X(\tau_0))^\mu u^\nu_{\sim c}-(x-X(\tau_0))^\nu u^\mu_{\sim c}}{[u_{\sim c} \cdot (x-X(\tau_0))]^3}$
(see below), the Lorentz transformation technique we have used is valid at all the
time-space points $x^\mu$ to obtain the expression of the Coulomb electric field
produced by the electron at rest from the expression of the electromagnetic
field produced by the electron in uniform motion at its highest speed
(which is arbitrarily close to the speed of light $v\sim c$).

From eq. (\ref{3}) we find
\bea
F'^{0i} = \frac{\partial x'_0}{\partial x^\alpha}\frac{\partial x'^i}{\partial x^\beta} F^{\alpha \beta}(x).
\label{c3a}
\eea
From eqs. (\ref{c3a}) and (\ref{eyc13}) we find
\bea
F'^{0i} = \gamma_{\sim c} \beta_{\alpha_{\sim c}} [ g_{i \delta} + \frac{\gamma_{\sim c} -1}{{\vec {\beta}}^2_{\sim c}} \beta_{j_{\sim c}} g_{j \delta} {\beta}^i_{\sim c} - \gamma_{\sim c} \beta^i_{\sim c} g_{0\delta}]
 F^{\alpha \delta}(x).
\label{c13a}
\eea
Since $F^{\mu \nu}(x)$ in eq. (\ref{c5ixx}) is neither (exactly) infinity nor (exactly) zero (see eqs. (\ref{inf}) and (\ref{nexpg}))
we find by using eq. (\ref{c5ixx}) in (\ref{c13a})
\bea
&&F'^{0i} =\frac{ec}{[u_{\sim c} \cdot (x-X(\tau_0))]^3}~\left[[ g^i_\delta + \frac{\gamma_{\sim c} -1}{{\vec {\beta}}^2_{\sim c}} \beta_{j_{\sim c}} g_{j \delta} {\beta}^i_{\sim c} - \gamma_{\sim c} \beta^i_{\sim c} g_{0\delta}] \right]\nonumber \\
&&~[u_{\sim c} \cdot (x-X(\tau_0)) u^\delta_{\sim c} - (x-X(\tau_0))^\delta c^2].
\label{c13b}
\eea
Since electron has non-zero mass we find from eq. (\ref{ghf})
that $F'^{0i} \neq 0$ and $F'^{0i} \neq \infty $ in eq. (\ref{c13b}). Hence we find from eq. (\ref{c13b})
\bea
&& F'^{0i}=\frac{ec}{[u_{\sim c} \cdot (x-X(\tau_0))]^3}~[ [u_{\sim c} \cdot (x-X(\tau_0)) u^i_{\sim c}- (x-X(\tau_0))^ic^2 ]+ [\frac{\gamma_{\sim c} -1}{{\vec {\beta}}^2_{\sim c}} \beta_{j_{\sim c}} {\beta}^i_{\sim c}] \nonumber \\
&&[u_{\sim c} \cdot (x-X(\tau_0)) u_{j_{\sim c}}- (x-X(\tau_0))_jc^2 ]- \gamma_{\sim c} \beta^i_{\sim c} [u_{\sim c} \cdot (x-X(\tau_0)) u_{0_{\sim c}} - (x-X(\tau_0))_0c^2]].\nonumber \\
\label{c13c}
\eea
Simplifying eq. (\ref{c13c}) we find
\bea
&& F'^{0i} =\frac{ec}{[u_{\sim c} \cdot (x-X(\tau_0))]^3}~[ [u_{\sim c} \cdot (x-X(\tau_0)) u^i_{\sim c}- (x-X(\tau_0))^i c^2]\nonumber \\
&&+ (\gamma_{\sim c} -1) {\beta}^i_{\sim c} [c ~u_{\sim c} \cdot (x- X(\tau_0)) \gamma_{\sim c} - \frac{{\vec \beta}_{\sim c} \cdot ({\vec x}-{\vec X}(\tau_0))c^2}{{\vec \beta}^2_{\sim c}} ]\nonumber \\
&&- \gamma_{\sim c} \beta^i_{\sim c} [c~u_{\sim c} \cdot (x-X(\tau_0)) \gamma_{\sim c} - (x_0-X_0(\tau_0))c^2]].
\label{c13d}
\eea
Using eqs. (\ref{eyc13}) and (\ref{eyttpr}) in (\ref{c13d}) we find
\bea
&& F'^{0i} =\frac{e}{c(x'_0-X'_0(\tau_0'))^3}~[ [(x'_0-X'_0(\tau_0') u^i_{\sim c}- (x-X(\tau_0))^i c ]\nonumber \\
&&+ c(\gamma_{\sim c} -1) {\beta}^i_{\sim c} [(x'_0- X'_0(\tau_0')) \gamma_{\sim c} - \frac{{\vec \beta}_{\sim c} \cdot ({\vec x}-{\vec X}(\tau_0))}{{\vec \beta}^2_{\sim c}} ]- c\gamma_{\sim c} \beta^i_{\sim c} [(x'_0-X'_0(\tau_0)) \gamma_{\sim c} - (x_0-X_0(\tau_0))]]. \nonumber \\
\label{c13e}
\eea
The retarded condition is given by
\bea
x'_0-X'_0(\tau_0) = |{\vec x}' - {\vec X}'(\tau_0')| =R'
\label{crtd}
\eea
By using eq. (\ref{crtd}) in (\ref{c13e}) we find
\bea
&& F'^{0i} =\frac{e}{R'^3}~\left[ R' \gamma_{\sim c} \beta^i_{\sim c}- Rn^i + (\gamma_{\sim c} -1) {\beta}^i_{\sim c} [R' \gamma_{\sim c} - R\frac{{\vec \beta}_{\sim c} \cdot {\vec n} }{{\vec \beta}^2_{\sim c}} ]- \gamma_{\sim c} \beta^i_{\sim c} [R' \gamma_{\sim c} - R]\right] \nonumber \\
\label{c13f}
\eea
where we have introduced the notation
\bea
{\vec R} = {\vec n} R = {\vec x} - {\vec X}(\tau_0).
\eea
From eq. (\ref{c13f}) we find
\bea
&& F'^{0i} =\frac{e R}{R'^3}~\left[- n^i + \frac{{\vec \beta}_{\sim c} \cdot {\vec n} }{{\vec \beta}^2_{\sim c}} \beta^i_{\sim c} + \gamma_{\sim c}  {\beta}^i_{\sim c} [1 - \frac{{\vec \beta}_{\sim c} \cdot {\vec n} }{{\vec \beta}^2_{\sim c}} ]\right]
\label{c13g}
\eea
which gives
\bea
&& F'^{0i} =-\frac{e }{R'^3}~\left[R^i -\frac{{\vec \beta}_{\sim c} \cdot {\vec R} }{{\vec \beta}^2_{\sim c}} \beta^i_{\sim c} - \gamma_{\sim c}  {\beta}^i_{\sim c} [R - \frac{{\vec \beta}_{\sim c} \cdot {\vec R} }{{\vec \beta}^2_{\sim c}} ]\right].
\label{c13h}
\eea
Eq. (\ref{c13h}) can be simplified to find
\bea
&& F'^{0i} =-\frac{e }{R'^3}~\left[R^i +(\gamma_{\sim c} -1) \frac{{\vec \beta}_{\sim c} \cdot {\vec R} }{{\vec \beta}^2_{\sim c}} \beta^i_{\sim c} - \gamma_{\sim c}  {\beta}^i_{\sim c} (x_0-X_0(\tau_0))\right].
\label{c13icv}
\eea
Hence we find from eqs. (\ref{crtd}) and (\ref{c13icv})
\bea
&& F'^{0i} =-\frac{e }{|{\vec x}' -{\vec X}'(\tau_0')|^3}~\left[(x^i -X^i(\tau_0)) +(\gamma_{\sim c} -1) \frac{{\vec \beta}_{\sim c} \cdot ({\vec x} -{\vec X}(\tau_0))}{{\vec \beta}^2_{\sim c}} \beta^i_{\sim c} - \gamma_{\sim c}  {\beta}^i_{\sim c} (x_0-X_0(\tau_0))\right].\nonumber \\
\label{c13i}
\eea

By using eqs. (\ref{eyc13}) and (\ref{eyttpr}) in eq. (\ref{c13i}) we find
\bea
&& F'^{0i} =-e\frac{x'^i -X'^i(\tau_0')}{|{\vec x}' -{\vec X}'(\tau_0')|^3}.
\label{c13j}
\eea
\\

Hence from eqs. (\ref{eyttpr}) and (\ref{c13j}) we find
\bea
F^{0i} =-e\frac{x^i -X^i(\tau_0)}{|{\vec x} -{\vec X}(\tau_0)|^3}
\label{c14k}
\eea
which is the exact expression of the Coulomb electric field produced by the electron at rest.

Hence we find that the exact expression of the Coulomb electric field in eq. (\ref{c14k}) produced
by the electron at rest can be obtained from the expression of (approximate) pure gauge electromagnetic field
$F^{\mu \nu}(x)$ from eq. (\ref{pgf}) [or from eq. (\ref{c5ixx})] produced by
the electron in uniform motion at its highest speed (which is arbitrarily close to the speed of
light $v \sim c$) by using Lorentz transformations.

Hence one finds that if one knows the expression of the (approximate) pure gauge potential produced by a point
charge of non-zero mass in uniform motion at its highest speed (which is arbitrarily close to the speed
of the light $v \sim c$), then one can obtain the exact expression of the potential produced by the same charge
at rest by using Lorentz transformations. This is consistent with the replacement given by eq. (\ref{my2})
which can be used to obtain eq. (\ref{colf}) from (\ref{pglw}) and eq. (\ref{c14k}) from (\ref{pgf}) [or
from eq. (\ref{c5ixx})].

\section{ Yang-Mills Color Current Density of the Quark }

Note that the Lienard-Wiechert potential (Maxwell potential) $A^\mu(x)$ in eq. (\ref{1yj1}) can be obtained from
the abelian current density
\bea
j^\mu(x) = e \int d\tau ~u^\mu(\tau)~ \delta^{(4)}(x-X(\tau))
\label{cm}
\eea
as given by eq. (\ref{mxcz}) of the point charge $e$ in Maxwell theory [see section III for details].
However, the non-abelian Yang-Mills potential $A^{\mu a}(x)$ in eq. (\ref{tpg1}) gives
the non-abelian Yang-Mills color current density of the quark
\bea
j^{\mu a}(x) =D_\nu [A]F^{\nu \mu a}(x)
\label{ccvn}
\eea
which contains infinite powers of $g$ [or infinite power of fundamental
time dependent color charge $q^a(\tau)$ of the quark] where $D_\mu^{ab}[A]$ is given by
eq. (\ref{dm}) and $F^{\mu \nu a}(x)$ is given by eq. (\ref{fmna}). This implies that
the non-abelian Yang-Mills color
current density $j^{\mu a}(x)$ of the quark in eq. (\ref{ccvn}) is not linearly proportional
to $g$ [and is not linearly proportional to fundamental
time dependent color charge $q^a(\tau)$ of the quark], {\it i.e.},
\bea
j^{\mu a}(x) \neq \int d\tau ~u^\mu(\tau) ~q^a(\tau) ~\delta^{(4)}(x-X(\tau)).
\label{cy}
\eea
Hence one finds that not only the non-abelian Yang-Mills potential $A^{\mu a}(x)$ in eq. (\ref{tpg1})
contains infinite powers of $g$ [or infinite powers of fundamental time dependent color charge
$q^a(\tau)$ of the quark] but also the non-abelian Yang-Mills color current density $j^{\mu a}(x)$
of the quark in eq. (\ref{ccvn}) or in eq. (\ref{1a}) contains infinite powers of $g$ [or infinite powers
of fundamental time dependent color charge $q^a(\tau)$ of the quark].

\end{document}